\documentclass[twocolumn,showpacs,preprintnumbers,amsmath,amssymb,prb]{revtex4}

\usepackage{graphicx}
\usepackage{dcolumn}
\usepackage{bm}
\begin{document}


\title{Taming the Rugged Landscape: Production, Reordering, and
Stabilization of Selected Cluster Inherent Structures in the
$\mathrm{X_{13-n}Y_n}$ System}

\author{Dubravko Sabo}
\author{J. D. Doll}
\affiliation{Department of Chemistry, \\ Brown University, \\ Providence,
RI 02912, USA}

\author{David L. Freeman}
\affiliation{Department of Chemistry, \\ University of Rhode Island,
\\ Kingston, RI 02881, USA}

\date{\today}

\begin{abstract}
We present studies of the potential energy landscape of selected binary
Lennard-Jones thirteen atom clusters. The effect of adding selected
impurity atoms to a homogeneous cluster is explored. We analyze the
energy landscapes of the studied systems using disconnectivity graphs.
The required inherent structures and transition states for the construction
of disconnectivity graphs are found by combination of conjugate gradient
and eigenvector-following methods. We show that it is possible to 
controllably induce new structures as well as reorder and stabilize
existing structures that are characteristic of higher-lying minima.
Moreover, it is shown that the selected structures can have experimentally
relevant lifetimes.
\end{abstract}


\maketitle

\section{Introduction} \label{sec:intro}

The science of chemistry is characterized by an interplay of reductionist
and constructionist themes. On the one hand, one seeks to reduce complex
systems to more understandable and more controllable components. On the
other, one strives to utilize these components to construct complex
assemblies to meet specific chemical, biological, and/or materials goals.
The core elements of such efforts have, in the past, been largely atomic
and/or molecular in nature.

Viewed from the above perspective, a central component of much ongoing
research has been the development of a variety of ``reductionist''
classical and quantum-mechanical tools. Common goals in the application
of these tools have been locating the minimum energy configuration for
the associated potential energy surface \cite{WALES00} and sampling
the relevant, low-lying local minima. \cite{FREEMAN00A,FREEMAN00B,JORD02A,
JORD03A}

Increasingly, both chemistry and materials science are entering an era
in which the fundamental components of the constructionist phase of the
problem often are themselves complex, preassembled objects. Sun,
\emph{et al.}, \cite{SUN00} for example, have shown that novel
magnetic materials can be prepared through the self-assembly of 
colloidal clusters. In related developments Lehmann and Scoles 
\cite{SCOLES98A} as well as Miller, \emph{et al.} \cite{REMILLER99A,
REMILLER99B,REMILLER99C} have demonstrated the effectiveness of superfluid
solvents in preparing unusual, metastable species. Sustained progress in
this emerging field of cluster assembled materials ultimately rests on
the ability to characterize and control a broad range of increasingly
complex nanoscale objects.

In a previous paper \cite{SABO03A}, we have examined a number of theoretical
issues of general concern with respect to 
predicting/characterizing/controlling the structure and dynamics of
cluster-based precursors. That work, in essence, seeks to invert the logic
of the minimization problem. That is, instead of searching for the 
minimum energy structures of specified energy landscapes, we strive instead
to reshape those landscapes and thereby to exercise control over selected
physical systems. In particular, we seek to stabilize and/or kinetically
trap conformers of the parent homogeneous system that are otherwise
either un- or metastable. Our hope is that such new structures might have
interesting physical properties (electronic, magnetic, optical, thermal,
etc.) and can be used as precursors in subsequent assembly procedures.
Our previous work \cite{SABO03A} has demonstrated that for small Lennard-Jones 
systems we could, using selected impurities, both alter the energy ordering 
of the stable core atom isomers and induce wholly new conformers not seen
in the original homogeneous species. We have also seen similar reordering
effects in our studies of molecular nitrogen adsorbates on nickel
clusters.\cite{FREEMAN04A}

In the present study, we wish to extend the results of our previous
investigations. First, we wish to consider a number of
larger clusters to demonstrate the general applicability of our efforts.
Second, we wish to demonstrate the resulting species can be made sufficiently
robust that they are of practical interest. To attain the second goal we
enhance our previous publication\cite{SABO03A} by determining
RRKM isomerization rate constants and lifetimes for the generated
isomers.

An important byproduct of the current work is the information
we garner concerning the structure of the potential energy surfaces
of the mixed cluster systems.  In the current work we
find that many of the disconnectivity graphs have double-funnel structures.
The thermodynamic properties of other cluster systems with double-funnel
structures\cite{WALES98A,FREEMAN00A} can be rich exhibiting
solid-like to solid-like phase change phenomena.
Motivated by these past studies and the disconnectivity graphs
investigated in the current work, in a companion paper\cite{SABO04B}
we investigate the
energy and heat capacity of some of the systems studied here.

The remainder of the paper is organized as follows. An outline of the
computational details of this work is presented in Sec~\ref{sec:comput}.
We discuss the methods utilized to find the inherent structures 
and transition states on the potential energy surface as well as the
method to estimate the inherent structures lifetimes. In Sec.~\ref{sec:numres}
we present the results that demonstrate ``proof of principle'' with respect
to the goals of the present studies. In Sec.~\ref{sec:conclude} we summarize
our results and speculate about likely future research directions.

\section{Computational Details} \label{sec:comput}

In this section, we describe the computational details of the
studies involving binary clusters of the form $\mathrm{X_{13-n}Y_n}$. Our
overall interest is to explore the extent to which we can
utilize the ``adatoms'' (i.e. the Y-system) to induce, reorder and
stabilize selected inherent structures in the ``core'' X-system. While
one can easily imagine applications involving more and more
complex components, we have found that these relatively simple, 
two-component clusters are a convenient starting point 
for an initial study of the issues we raise.\cite{SABO03A}

The total potential energy, $V_{tot}$, of a cluster consisting of
$N$ particles is modeled as a pairwise sum of Lennard-Jones interactions

\begin{equation}
\label{2.1}
V_{tot} = \sum_{i<j}^{N} v_{ij}(r_{ij}),
\end{equation}
where the pair interaction as a function of the distance between
particles $i$ and $j$, $r_{ij}$, is given by

\begin{equation}
\label{2.2}
v_{ij}(r_{ij}) = 4\epsilon_{ij}\  
      [( \frac{\sigma_{ij}}{r_{ij}})^{12}
     -( \frac{\sigma_{ij}}{r_{ij}})^{6}] .
\end{equation}
In Eq. (\ref{2.2}) the constants $\epsilon_{ij}$ and $\sigma_{ij}$ are
the energy and length-scale parameters for the interaction of
particles $i$ and $j$.

For a binary system, both the ``like'' (X-X, Y-Y) as well as the 
``unlike'' (X-Y) interactions have to be specified. With an eye
toward studying trends in the results as opposed to results for 
particular physical systems, it is convenient to reduce the number
of free parameters. To do so, we shall assume in the present study
that the ``unlike'' Lennard-Jones values are obtained from the ``like''
Lennard-Jones parameters via usual combination rules \cite{COMBINE}
\begin{equation}
\label{2.3}
\sigma_{XY}= \frac{1}{2}(\sigma_{XX}+\sigma_{YY}),
\end{equation}
\begin{equation}
\label{2.4}
\epsilon_{XY}= \sqrt{\epsilon_{XX}\epsilon_{YY}} .
\end{equation}
Furthermore, we note that with the mixed Lennard-Jones parameters
specified as in Eqs.(\ref{2.3}) and (\ref{2.4}), the inherent structure
topography of the ``reduced'' potential energy surface of the binary
system (i.e. $V_{tot}/\epsilon_{XX}$) is a function of only two
parameters, ($\sigma, \epsilon$), the ratios of the corresponding
adatom/core length and energy parameters
\begin{equation}
\label{2.5}
\sigma= \sigma_{YY}/\sigma_{XX},
\end{equation}
\begin{equation}
\label{2.6}
\epsilon= \epsilon_{YY}/\epsilon_{XX} .
\end{equation}

If necessary for a discussion of a specific physical system, the
absolute bond lengths, energies, activation energies, etc. can be
obtained from the corresponding ``reduced'' results by a simple
rescaling with the appropriate core-system Lennard-Jones parameters.

\subsection{Stationary points and the disconnectivity graph}

The computational task in our study is thus one of exploring and
characterizing the (reduced) potential energy surface of our
binary cluster systems as a function of the number of (core, adatom)
particles, ($\mathrm{n,m}$), and for given $(\sigma,\epsilon)$ ratios. In
typical applications the lowest $\mathrm N_{IS}$ inherent structures and the
associated disconnectivity graphs are determined. For the applications
reported here, $\mathrm N_{IS}$ is generally of the order of several 
thousands or more. The inherent structures are found either via
conjugate gradient methods\cite{PRESTEU} starting from randomly
chosen initial configurations, or by more systematic surface exploration
methods.\cite{WALES99B,JORD93B}
In all cases, the inherent structures that are located are 
confirmed to be stable minima via a standard Hessian analysis. To
reduce the chance we miss particular local or global minima, we monitor
the number of times individual inherent structures are found and
demand that each of the $\mathrm N_{IS}$ inherent structures be located a
minimum number of times (at least 10) before we terminate our search.
Once we are satisfied we have located the relevant inherent structures,
transitions states linking these stable minima are obtained using the
eigenvector following methods outlined by Cerjan and Miller \cite{MILLER81}
and further developed by Simons {\it et al.} \cite{SIMONS83,SIMONS85,SIMONS90},
Jordan {\it et al.} \cite{JORD93B} and Wales \cite{WALES94}.
Finally, with the given inherent structures and transition states,
we perform a disconnectivity analysis.\cite{KARPLUS97,WALES99A}

\subsection{Rate constants and lifetimes of the inherent structures} 

From the known inherent structures and the transition states that 
connect/separate them, we estimate rates for transitions between 
neighboring inherent structures. The rate constants allow us
to calculate the average amount of time the system will spend in a
given inherent structure, i.e. the lifetime of an inherent structure.

There are variety of methods available to estimate the rate constants
(see Ref.\onlinecite{BERRY95} and references therein). We utilize
the harmonic approximation to the Rice-Ramsperger-Kassel-Marcus
(RRKM) method. It has been found that this method gives good
estimates of rates for isomerization of clusters\cite{BERRY92}.
The rate constant, $k_{ij}$, for transition leading from
inherent structure $j$ to inherent structure $i$ is given as a sum over all
transition states connecting inherent structures $i$ and $j$ \cite{WALES00,
CALVO03}
\begin{equation}
\label{2.7}
k_{ij}=\sum_{\alpha}k_{j}^{\alpha}.
\end{equation}
$k_{j}^{\alpha}$ is given by 
\begin{equation}
\label{2.8}
k_{j}^{\alpha}=\frac{h_{j}^{\alpha} \prod_{l=1}^{3N-6}\nu_{l,j}^{IS}}
{h_{j} \prod_{l=1}^{3N-7}\nu_{l,j}^{TS,\alpha}}
e^{-\Delta \phi_j/k_BT}
\end{equation}
where $\Delta \phi_j$ is
\begin{equation}
\label{2.9}
\Delta \phi_j= \mathrm{E_j^{TS,\alpha}-E_j^{IS}},
\end{equation}
$\mathrm{E_j^{TS,\alpha}}$ is the energy of the transition state, 
$\mathrm{E_j^{IS}}$
is the energy of the inherent structure, and $\nu_{l,j}$ are the
corresponding normal mode frequencies. $N$ is the number of particles
in the cluster while $h_j$ and $h_j^{\alpha}$ are the
order of the point group of inherent structure $j$ and transition state
$\alpha$, respectively. Since we are interested in the qualitative
estimates of the rate constants (lifetimes) we neglect the order of the
point group of inherent structure and transition state from our
calculations. We estimate that the error arising from this approximation
is a factor between 60 and 1 based on the reasoning that follows: 
The global minimum of the systems we study
is icosahedral in nature which means that the order of its point group is no
larger than $h_j$=60. The transition state that connects the global minimum
with a higher lying inherent structure has a lower symmetry than the global
minimum and therefore smaller order of the point group than the one
associated with the point group of the global minimum. As is made evident
below, the factor of 60 is unimportant to the determination of the order of
magnitude estimates of isomer lifetimes that are of interest to us
in the current context.  We would like
to point out that the numerical value of $\epsilon_{XX}$ and
$\sigma_{XX}$ used to calculate the rate constants are 119.8 K
and 3.405 {\AA}, respectively.

\section{Numerical Results} \label{sec:numres}

In the present Section, we illustrate the general themes we
introduced in Section~\ref{sec:intro} . 
We demonstrate that we can accomplish
three basic objectives. Specifically, we show that by adding selected
``impurity'' atoms to bare ``core'' systems, we can:

1. induce new ``core structures'' 

2. reorder the energies of existing core inherent structures, and 

3. stabilize selected inherent structures by controlling the activation 
energies that determine their isomerization kinetics.

For purposes of illustration, we examine numerical results for three,
thirteen atom Lennard-Jones systems involving ten, eleven and twelve core 
atoms, systems well-known from previous studies \cite{FAKEN01} to have
64, 170 and 515  energetically distinct
inherent structures, respectively. These systems have been chosen because
they build upon simple ten, eleven and twelve-atom cores and because 
total systems have thirteen atoms, a magic number for icosahedral growth
in homogeneous systems.

\subsection{$\mathrm{X_{12}Y_1}$}

\begin{figure}[!htbp] \centering
  \begin{tabular}{@{}cc@{}}
    \includegraphics[width=3.4cm,clip=true]{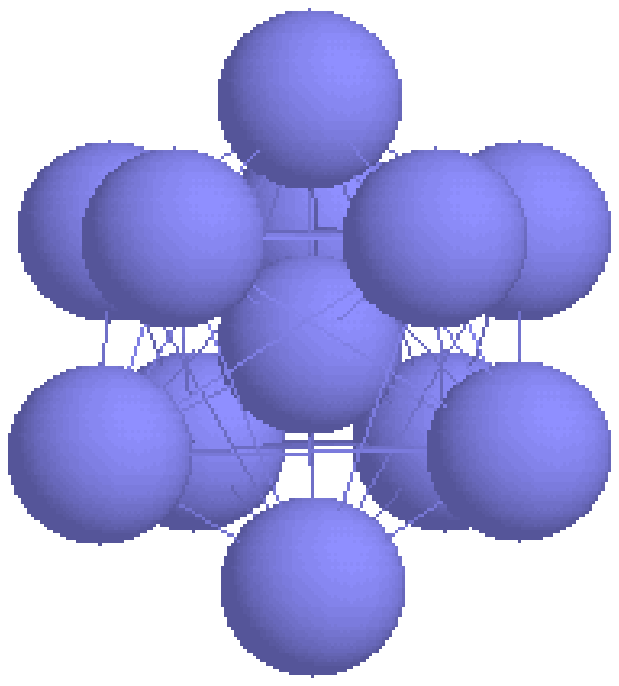} &
    \includegraphics[width=3.4cm,clip=true]{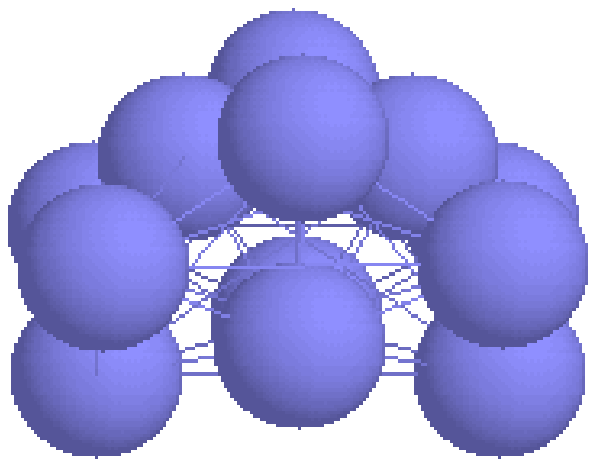} \\
    (a) & (b) \\
    \includegraphics[width=3.6cm,clip=true]{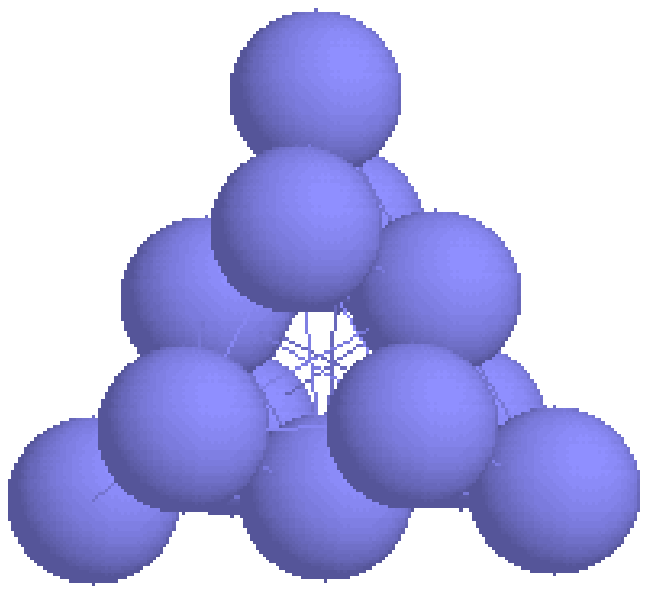} &
    \includegraphics[width=3.4cm,clip=true]{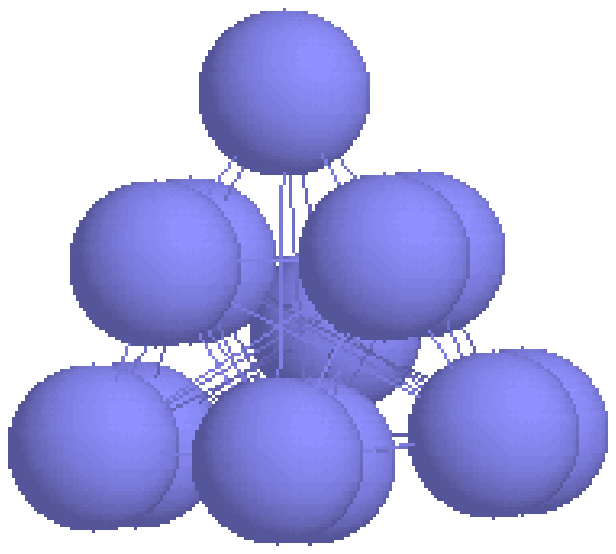} \\
    (c) & (d) \\
    \includegraphics[width=3.8cm,clip=true]{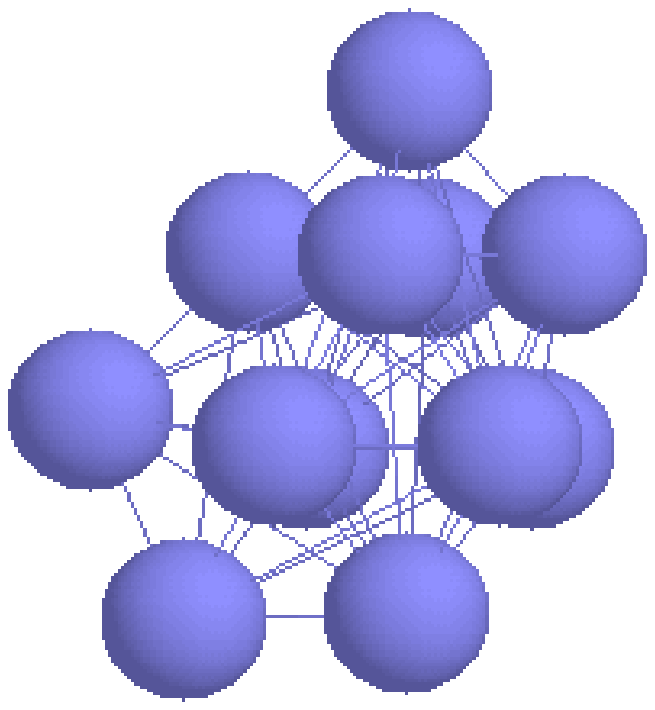} &
    \includegraphics[width=3.2cm,clip=true]{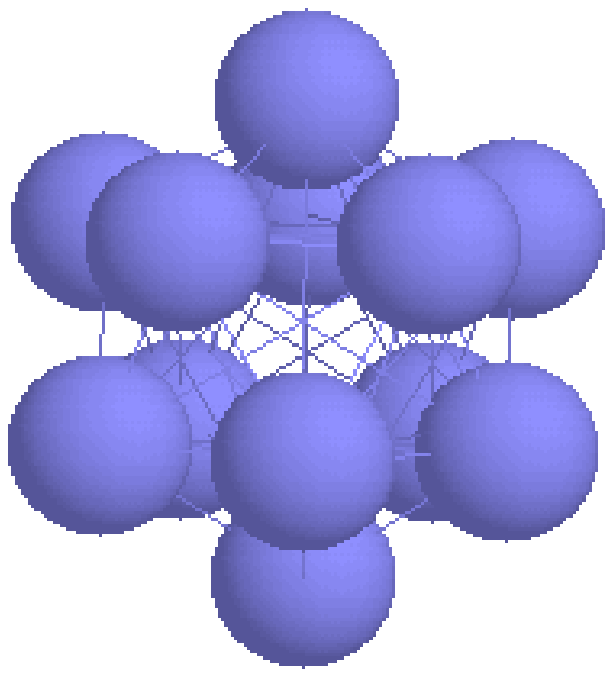} \\
    (e) & (f) \\
        \end{tabular}
\caption{\label{fig:core12} The selected stable inherent structure for X$_{12}$
LJ cluster. Their energies (in units of the LJ well depth) are: (a) -37.968,
(b) -36.347, (c) -36.209, (d) -35.461, (e) -34.447, (f) -33.598.}
\end{figure}

\begin{figure}
\includegraphics[clip=true,width=8.5cm]{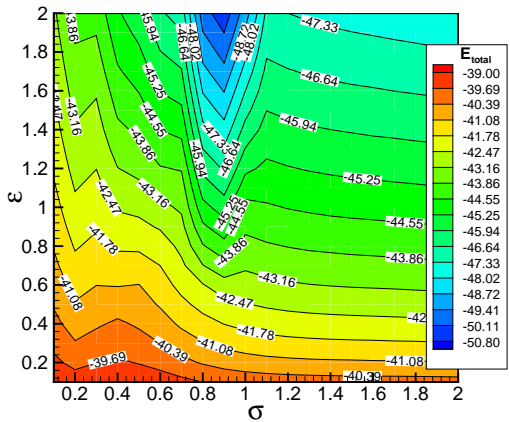}
\caption{\label{fig:totPESX12Y1}
$\mathrm E_{tot}(\sigma,\epsilon)$ (c.f.Eqs.\ref{2.1}, \ref{2.5} and \ref{2.6})
for the $\mathrm{X_{12}Y_1}$ system. Note the relative lack of structure
in the ($\sigma$,$\epsilon$) variation of the total cluster energy.}
\end{figure}

We first consider binary clusters of the type $\mathrm{X_{12}Y_1}$. 
The selected inherent structures and their associated energies
for $\mathrm{X_{12}}$ core system are illustrated in Fig.~\ref{fig:core12}.
The inherent structure labeled by (a) is the global minimum while
all others are the higher lying inherent structures.
Here one
impurity atom Y is added to the parent, twelve-atom X core. 
Using the
techniques of Section~\ref{sec:comput}, we determine the lowest several
inherent structures for a range of ($\sigma$,$\epsilon$) [c.f.
Eq.(\ref{2.5}) and Eq.(\ref{2.6})]. As can be seen from 
Fig.~\ref{fig:totPESX12Y1}, the total potential energy [Eq.(\ref{2.1})]
of the lowest inherent structure for the $\mathrm{X_{12}Y_1}$
system shows no appreciable structure as a function of the
($\sigma$,$\epsilon$) parameters.

\begin{figure}[!htbp]\centering
\includegraphics[clip=true,width=8.5cm]{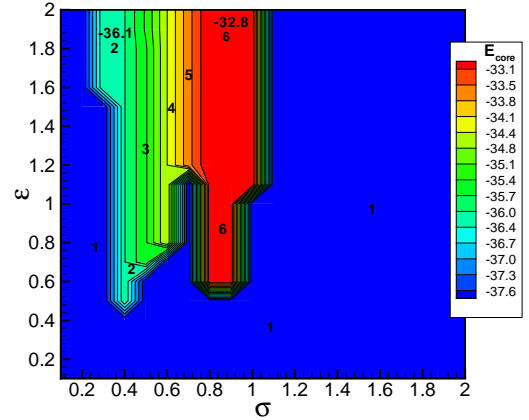}
\caption{\label{fig:corePESX12Y1}
$\mathrm E_{core}(\sigma,\epsilon)$ for the $\mathrm{X_{12}Y_1}$ system.
Here the ``core'' energy is defined as that portion of the potential energy
arising from only the core-core atom interactions. Unlike the total energy,
the $\mathrm(\sigma,\epsilon)$ variation of the core cluster energy exhibits
relatively well-defined regions. The labels of each of these regions in
the figure correspond to the distinct core structures shown in
Fig.~\ref{fig:coreEX12Y1}.}
\end{figure}

On the other hand, it can be seen in Fig.~\ref{fig:corePESX12Y1}
that the core potential energy, defined as the
potential energy of interaction for only the core X-atoms, of the minimum
(total) energy cluster clearly breaks into extended regions, each
corresponding to a well-defined core structure. We would like to point out
that each region in Fig.~\ref{fig:corePESX12Y1} contains the same ``kind''
of core structure but their core energies are slightly different. We have
chosen a single ``average'' core energy value to represent all energies in the
corresponding domain for plotting convenience.

The distinct core structures, shown in Fig.~\ref{fig:corePESX12Y1}, have
been identified by examining their core energies (E$_{core}$) and their
principal moments of inertia. For each structure a triplet of
values (E$_{core}$, I$_2$, I$_3$) has been associated, where I$_2$ and
I$_3$ are the moments of inertia about the principal axes 2 and 3,
respectively. We have defined I$_2$ and I$_3$ in the following way:
I$_2$=I$^{'}_2$/I$^{'}_1$, I$_3$=I$^{'}_3$/I$^{'}_1$ where I$^{'}_1$,
I$^{'}_2$ and I$^{'}_3$ are the principal moments of inertia obtained
by diagonalizing the inertia tensor of the system. If the triplet of
values has not been sufficient to identify a core structure then we
have examined the structure visually.

\begin{figure}[!htbp] \centering
  \begin{tabular}{@{}cc@{}}
    \includegraphics[width=3.4cm,clip=true]{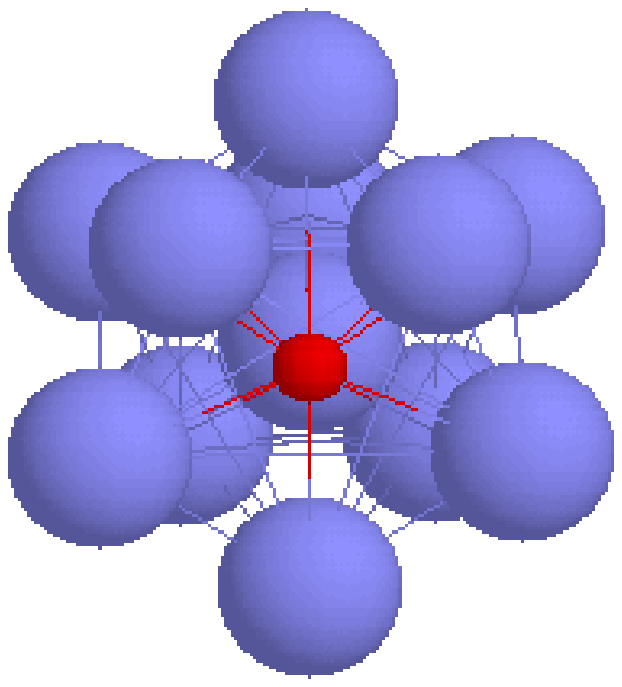} &
    \includegraphics[width=3.8cm,clip=true]{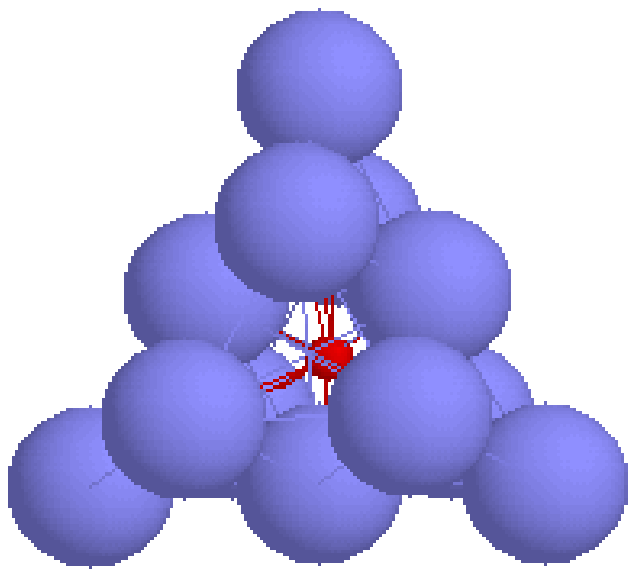} \\
    (4.1) & (4.2) \\
    \includegraphics[width=3.6cm,clip=true]{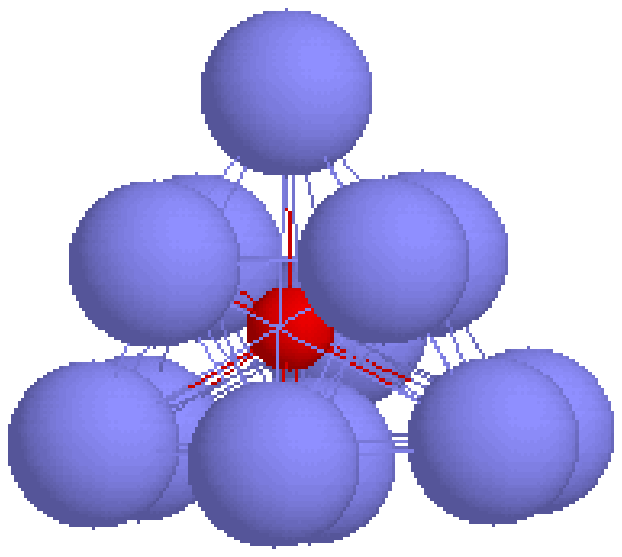} &
    \includegraphics[width=4.0cm,clip=true]{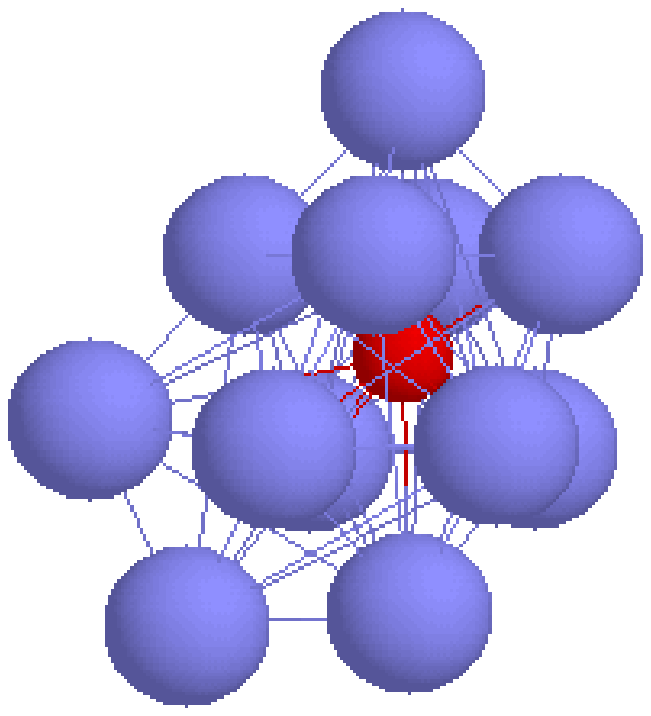} \\
    (4.3) & (4.4) \\
    \includegraphics[width=3.4cm,clip=true]{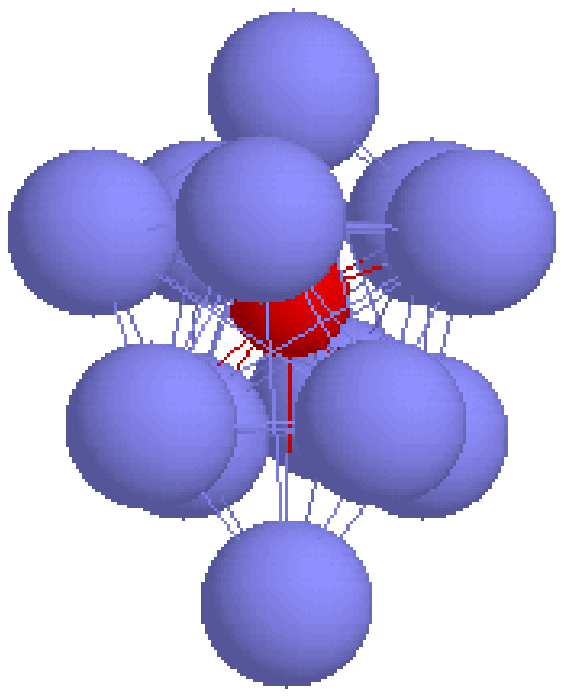} &
    \includegraphics[width=3.4cm,clip=true]{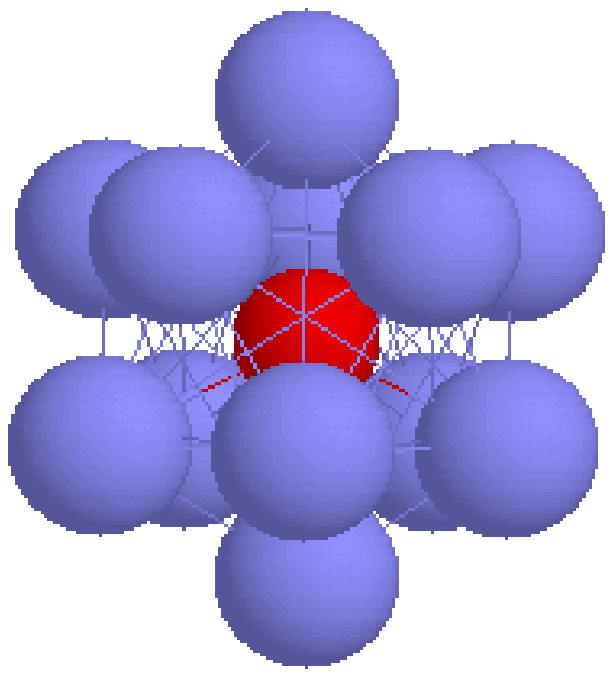} \\
    (4.5) & (4.6) \\
        \end{tabular}
\caption{\label{fig:coreEX12Y1}
Plots of $\mathrm{X_{12}Y_1}$ structures for selected
$\mathrm(\sigma,\epsilon)$ values. The decimal number for each figure 
denotes the corresponding $\mathrm(\sigma,\epsilon)$ domain in
Fig.~\ref{fig:corePESX12Y1}.}
\end{figure}

Selected cluster structures
illustrating the core arrangements corresponding to various
($\sigma$,$\epsilon$) values are shown in Fig.~\ref{fig:coreEX12Y1}.
It can be seen from Figs.~\ref{fig:corePESX12Y1} and \ref{fig:coreEX12Y1}
that the $\mathrm{X_{12}Y_1}$ cluster exhibits core X-atom structures 
that are higher lying minima in the parent
$\mathrm{X_{12}}$ system. The core structures labeled by (4.2), (4.3), 
(4.4), and (4.6) can be recognized as structures labeled by (c), (d), (e),
and (f) in Fig.~\ref{fig:core12}, respectively. This illustrates that a
suitable choice of the $\mathrm(\sigma,\epsilon)$ parameters can 
controllably reorder the energies of existing core inherent structures.
The structure labeled by (4.5) shows a newly induced core geometry not present
as a stable minimum in the bare cluster. These two results demonstrate
that we can accomplish objectives (1) and (2) stated earlier. 
\begin{figure*}
  \begin{tabular}{@{}cc@{}}
    \includegraphics[width=8.5cm,clip=true]{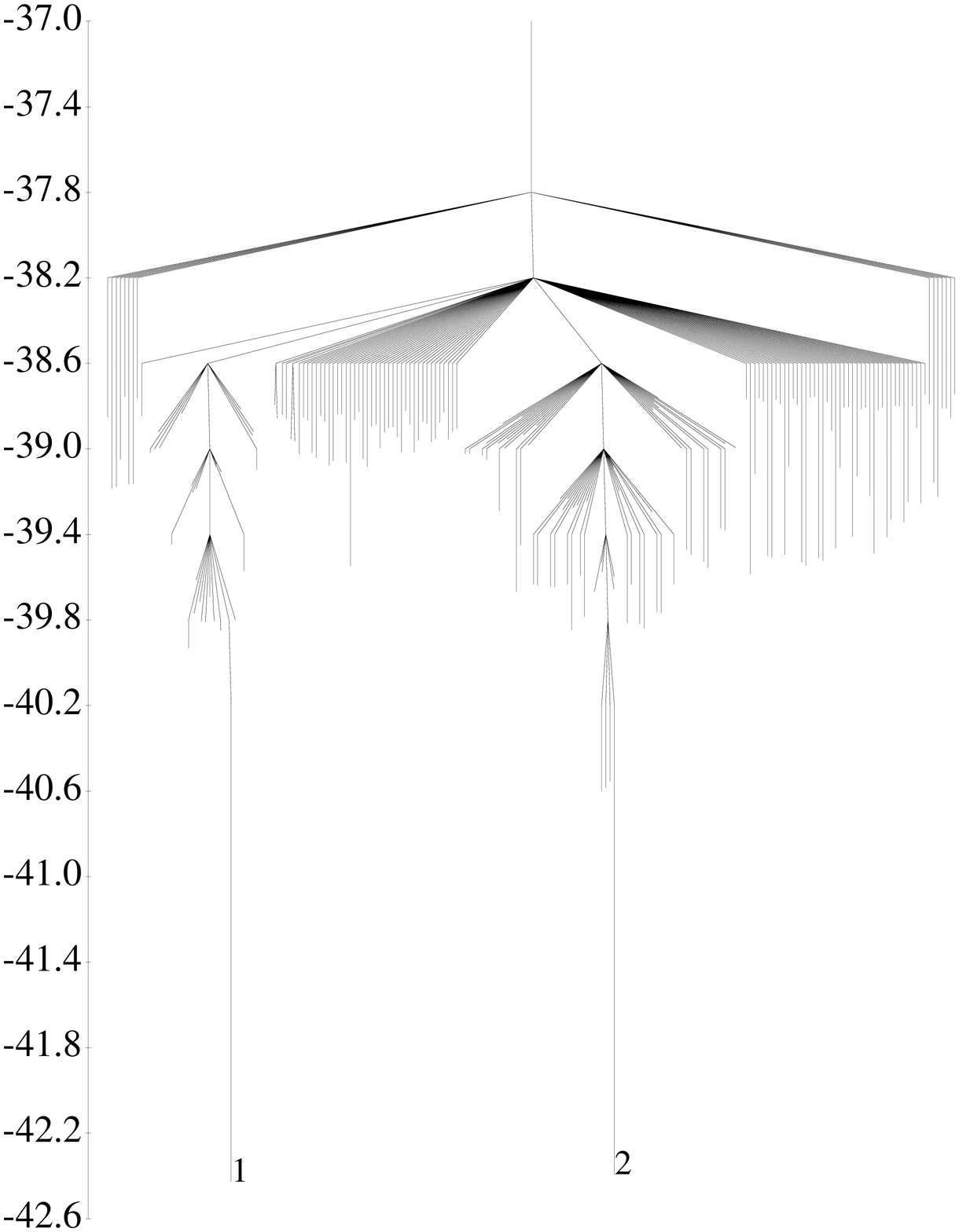} &
    \includegraphics[width=8.5cm,clip=true]{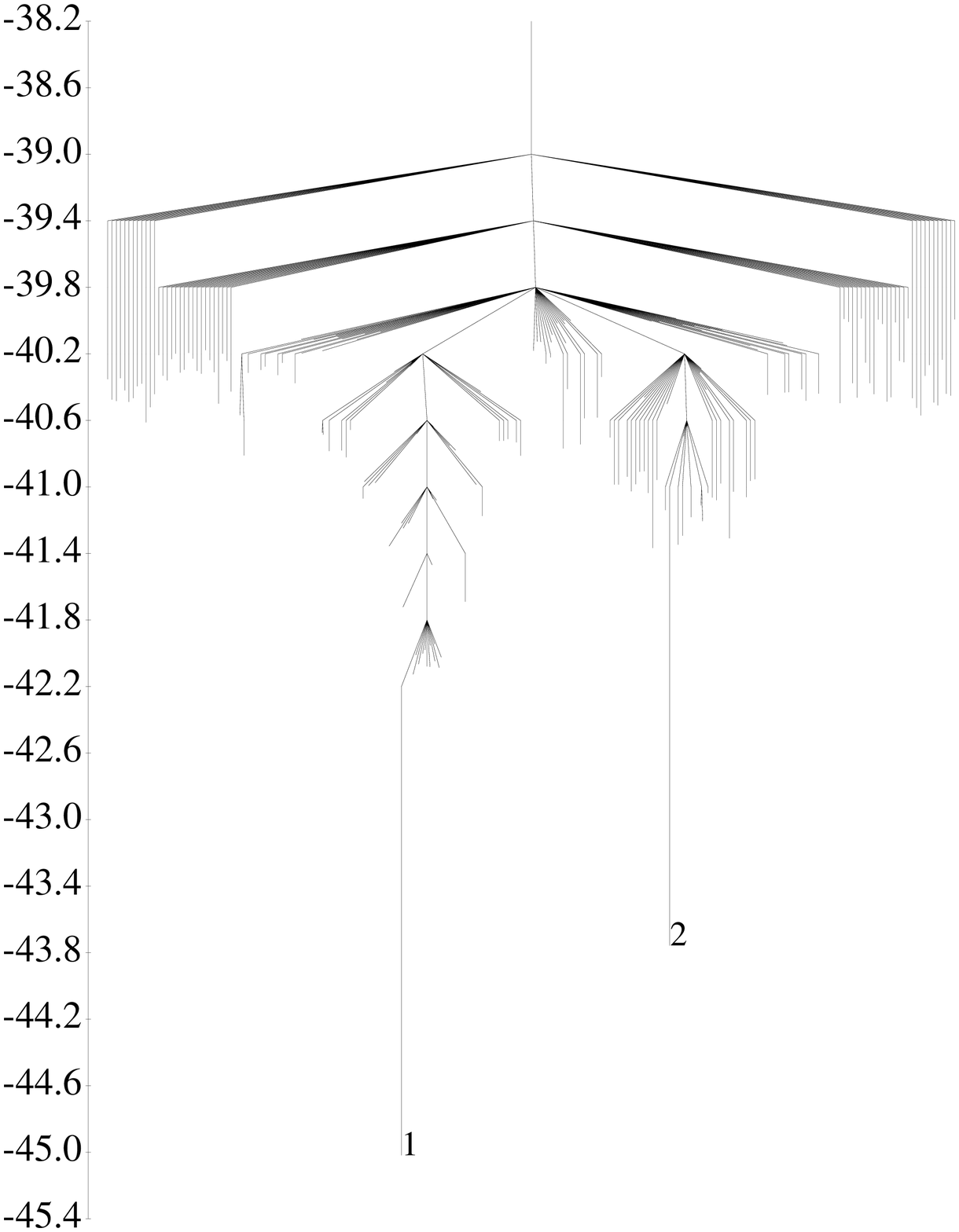} \\
    (a) & (b) \\
    \includegraphics[width=8.5cm,clip=true]{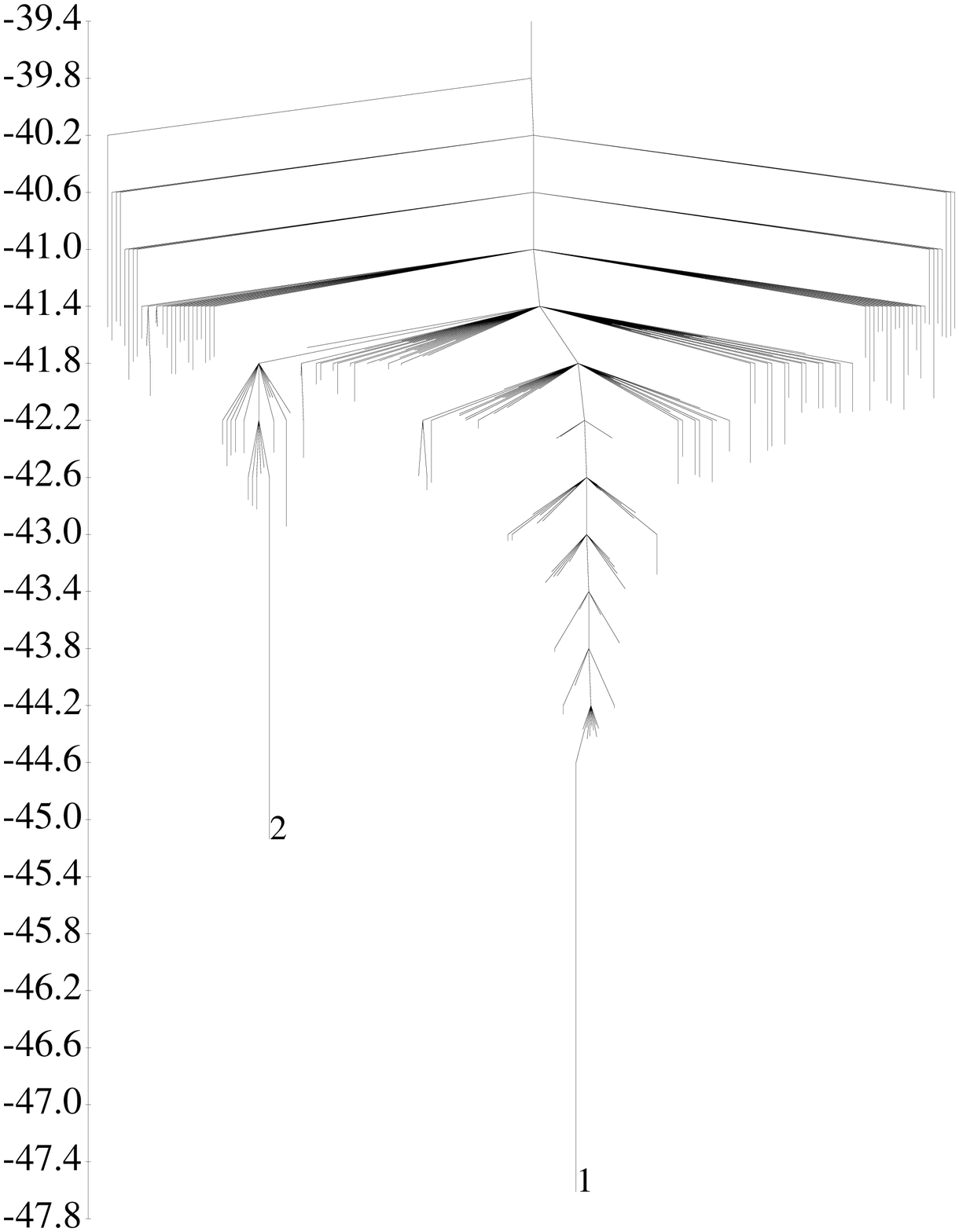} &
    \includegraphics[width=8.5cm,clip=true]{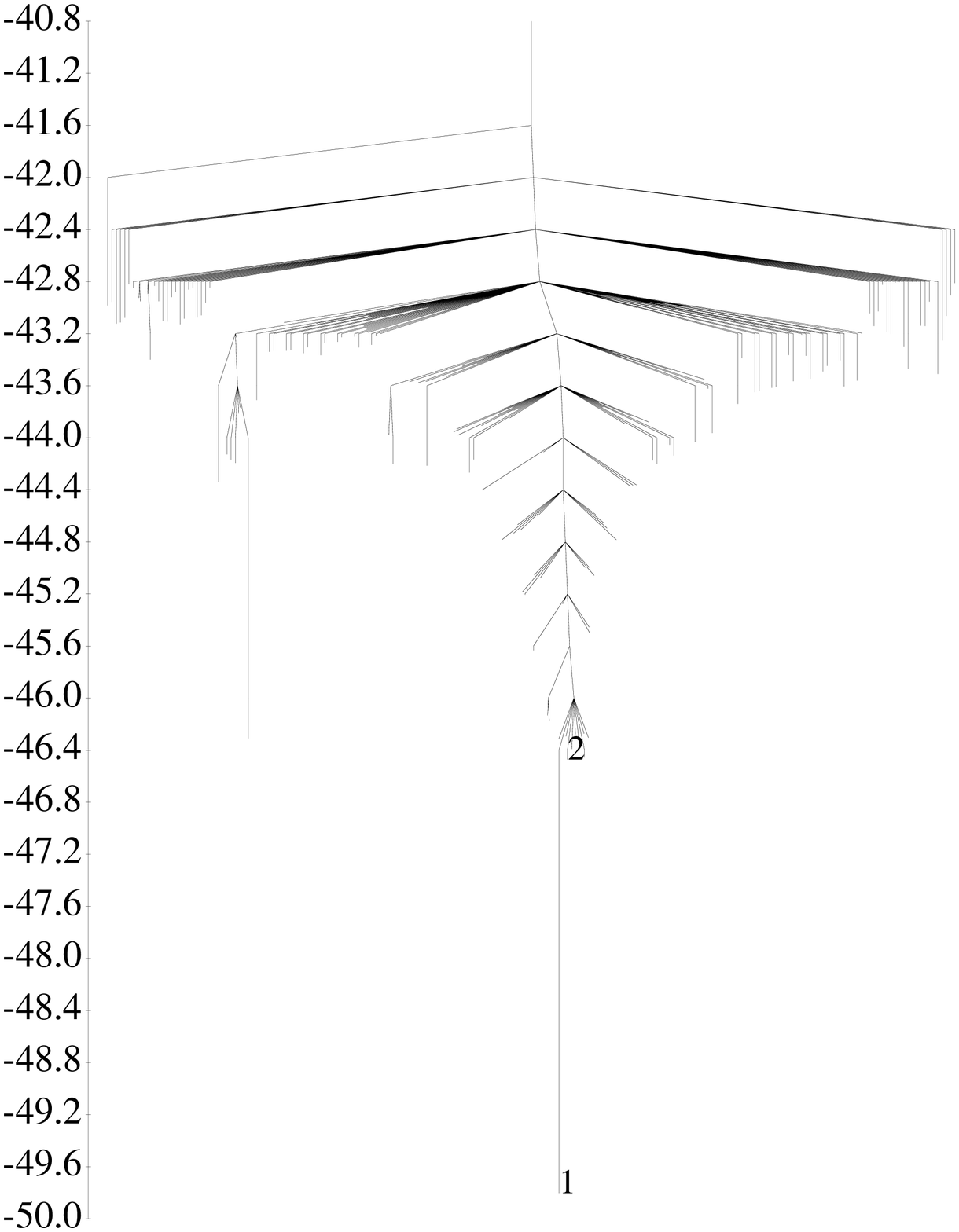} \\
    (c) & (d) \\
        \end{tabular}
\caption{\label{fig:DisConX12Y1}
Disconnectivity graph for $\mathrm{X_{12}Y_1}$$\mathrm(\sigma,\epsilon)$
values demonstrating that we can control
barriers for the selected inherent structures. The energy scale is in
units of $\epsilon_{XX}$. The $\mathrm(\sigma,\epsilon)$ values for
panels (a--d) are (0.8,0.6), (0.8,1.0), (0.8,1.5), and (0.8,2.0),
respectively. Only branches leading to the 200 lowest-energy minima are
shown.}
\end{figure*}

Figure ~\ref{fig:DisConX12Y1} represents the $\mathrm{X_{12}Y_1}$
cluster at four points in Fig.~\ref{fig:corePESX12Y1} defined by the
$\mathrm(\sigma,\epsilon)$ coordinates (0.8,0.6), (0.8,1.0), (0.8,1.5),
and (0.8,2.0). Here the pairs of coordinates correspond to (a), (b), (c),
and (d) of Fig.~\ref{fig:DisConX12Y1}, respectively. In other words, we
keep value of $\sigma$=0.8 fixed, while increasing the value of $\epsilon$.
The number of inherent structures available to the $\mathrm{X_{12}Y_1}$
cluster varies from at least 4153 in Fig.~\ref{fig:DisConX12Y1}.a to
at least 3641 in Fig.~\ref{fig:DisConX12Y1}.d.
Since we are primarily interested in energetically low-lying inherent
structures we show only the lowest 200 inherent structures. The global minimum
of each system is labeled by the number 1 and contains as a recognizable
component the core structure labeled by (4.6) (see Fig.~\ref{fig:coreEX12Y1}).
In Fig.~\ref{fig:DisConX12Y1}.a the core structure (4.6) is connected 
to a group of 12 inherent structures by pathways whose energies do not
exceed -39.4 (in units of $\epsilon_{XX}$). The energies of these 12
inherent structures are very close to each other and their values are
from lowest -39.810 to highest -39.614. Their corresponding core structures
are different from the core structure associated with the global minimum.
This implies that barriers that connect the global minimum with the inherent
structures in the group are ``relevant'' barriers. The relevant
barriers as those that connect inherent structures that contain different
core structures.
By examining the disconnectivity graph on a finer energy scale we find
that inherent structure 1 is connected, by the
lowest isomerization barrier, to inherent structure 16.
The numerical value of the lowest isomerization barrier 
is $\Delta$E$_{1,16}$=2.623$\epsilon_{XX}$.
Figures.~\ref{fig:DisConX12Y1}.b and
\ref{fig:DisConX12Y1}.c show that increasing the value of $\epsilon$
increases the isomerization barriers that connect inherent structure 1
with a group of 12 and 9 inherent structures, respectively. Similarly
to system (a), the inherent structures associated with both groups
contain core structures that are different from the core structure
associated with inherent structure 1. 
For system (b) the lowest isomerization barrier that connects
inherent structure 4 with inherent structure 1 has a
numerical value $\Delta$E$_{1,4}$=2.960$\epsilon_{XX}$ while for 
system (c) the lowest isomerization barrier connects inherent structure 1
with inherent structure 6 and 
its value is $\Delta$E$_{1,6}$=3.269$\epsilon_{XX}$. 
In Fig.~\ref{fig:DisConX12Y1}.d inherent structure 1 is connected to
a group of 9 inherent structures. The lowest isomerization barrier 
is $\Delta$E$_{1,3}$=3.509$\epsilon_{XX}$ connecting inherent structure
1 with inherent structure 3. The double-funnel structure of the
disconnectivity graphs are evident especially in
Fig.~\ref{fig:DisConX12Y1}.a and Fig.~\ref{fig:DisConX12Y1}.b
where the minima that define the two separated basins are so close 
in energy. The double-funnel structure of the potential energy
surface is reflected in the classical heat capacity as discussed in the
companion paper.\cite{SABO04B} Below we find similar double-funnel structures
for $\mathrm{X_{11}Y_2}$.

We estimate the rate constants (lifetimes) for four temperatures,
5, 10, 100, and 300 K, as a function of the height of the
isomerization barriers ($\epsilon$). At the low temperatures (5 and 10 K)
the studied systems become extremely stable. By increasing the isomerization
barrier between the global minimum and the first higher lying inherent 
structure, from $\Delta$E$_{1,16}$=2.623$\epsilon_{XX}$ 
(see Fig.~\ref{fig:DisConX12Y1}.a) to 
$\Delta$E$_{1,6}$=3.269$\epsilon_{XX}$ 
(see Fig.~\ref{fig:DisConX12Y1}.d) the lifetime increases by
nine and four orders of magnitude in the case of 5 and 10 K, respectively.
To be more specific, at 10 K, the lifetime increases from the order of
seconds to the order of days. 
This is illustrated in Fig.~\ref{fig:lifetimes}.a.

As illustrated in Fig.~\ref{fig:DisConX12Y1} and Fig.~\ref{fig:lifetimes}.a
the barriers that determine the isomerization
kinetics are sensitive to the ($\sigma$,$\epsilon$) values and can thus
be at least partially controlled. Therefore, we have created selected
structures that have experimentally relevant lifetimes.
These results are specific demonstrations
of goal (3) stated earlier.

\subsection{$\mathrm{X_{11}Y_2}$}

\begin{figure}[!htbp] \centering
  \begin{tabular}{@{}cc@{}}
    \includegraphics[width=3.6cm,clip=true]{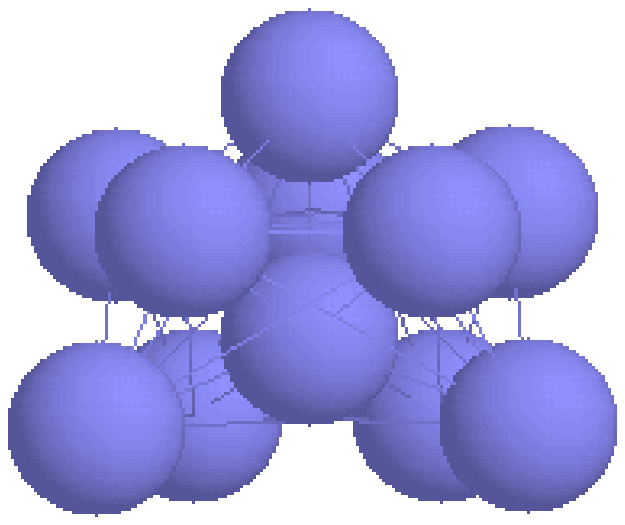} &
    \includegraphics[width=3.6cm,clip=true]{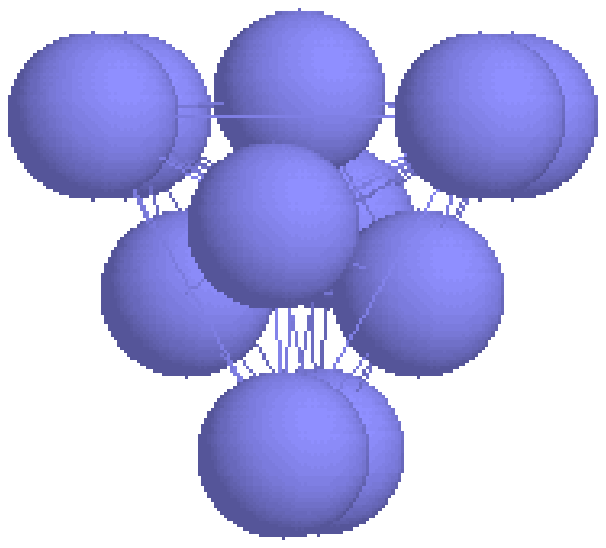} \\
    (a) & (b) \\
    \includegraphics[width=3.6cm,clip=true]{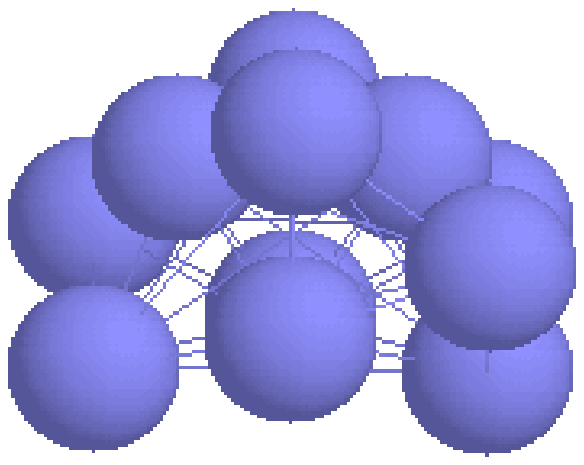} &
    \includegraphics[width=3.6cm,clip=true]{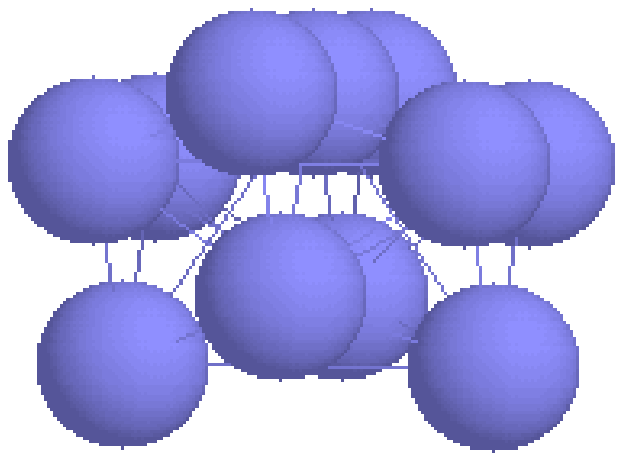} \\
    (c) & (d) \\
    \includegraphics[width=3.6cm,clip=true]{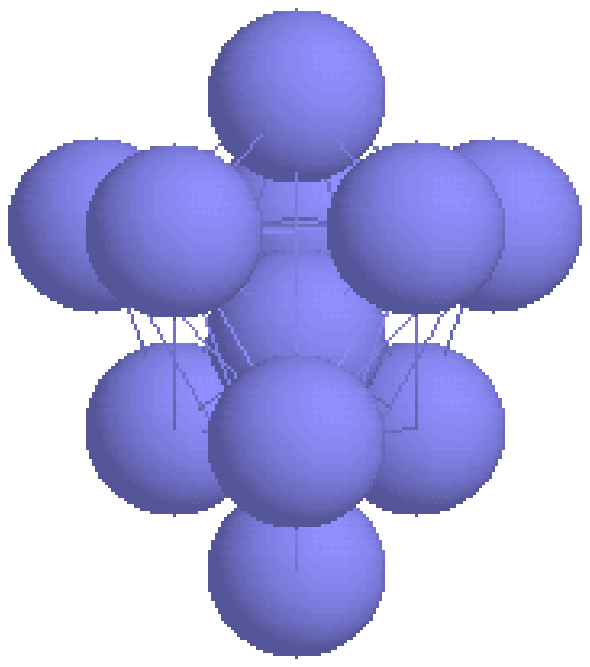} &
    \includegraphics[width=3.6cm,clip=true]{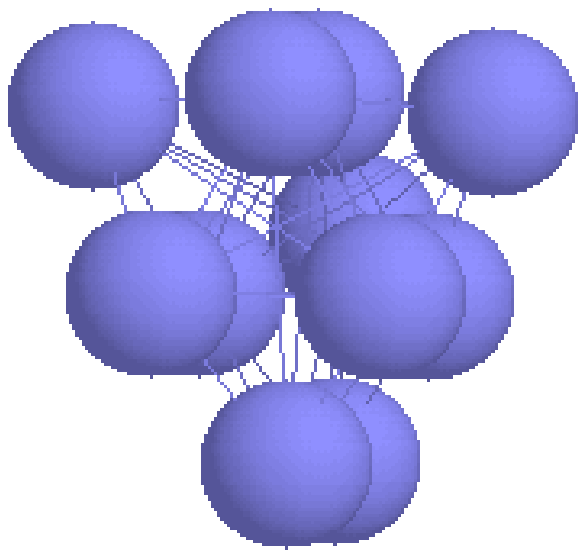} \\
    (e) & (f) \\
        \end{tabular}
\caption{\label{fig:core11} The selected stable inherent structure for 
$\mathrm{X_{11}}$ LJ cluster. Their energies (in units of the LJ well depth) 
are: (a) -31.766, (b) -31.9152, (c) -31.9146, (d) -31.775, (e) -31.615,
(f) -31.036.}
\end{figure}

\begin{figure}
\includegraphics[clip=true,width=8.5cm]{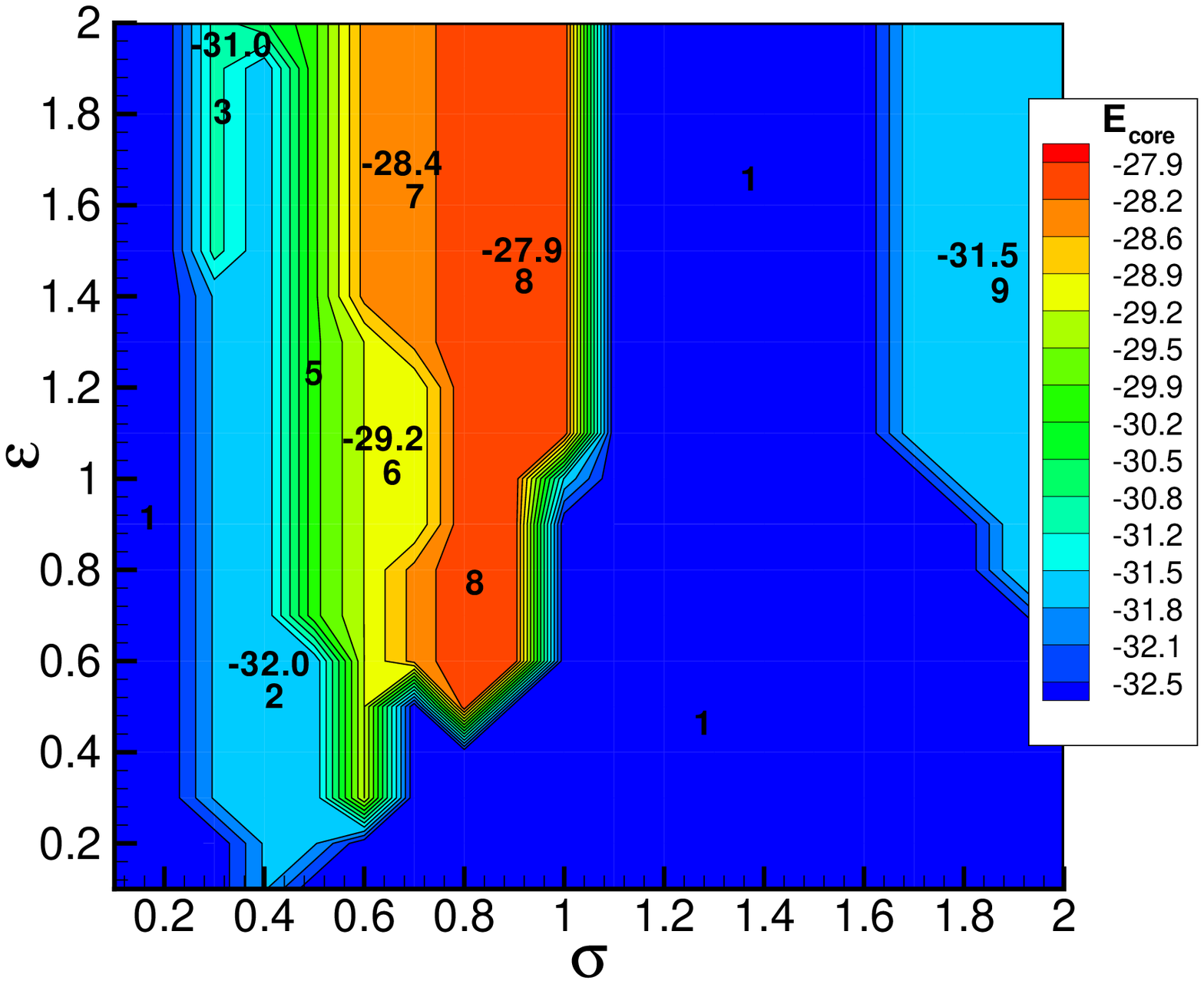}
\caption{\label{fig:corePESX11Y2}
$\mathrm E_{core}(\sigma,\epsilon)$ for $\mathrm{X_{11}Y_2}$.
Format for the plot is
the same as in Fig.~\ref{fig:corePESX12Y1}.}
\end{figure}

\begin{figure}[!htbp] \centering
  \begin{tabular}{@{}cc@{}}
    \includegraphics[width=3.8cm,clip=true]{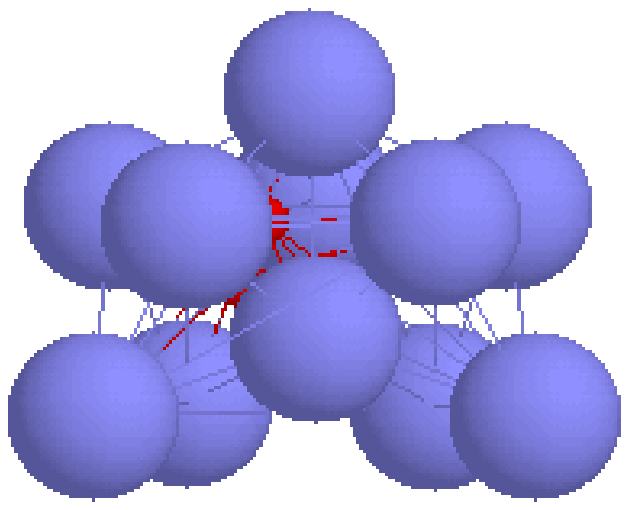} &
    \includegraphics[width=3.8cm,clip=true]{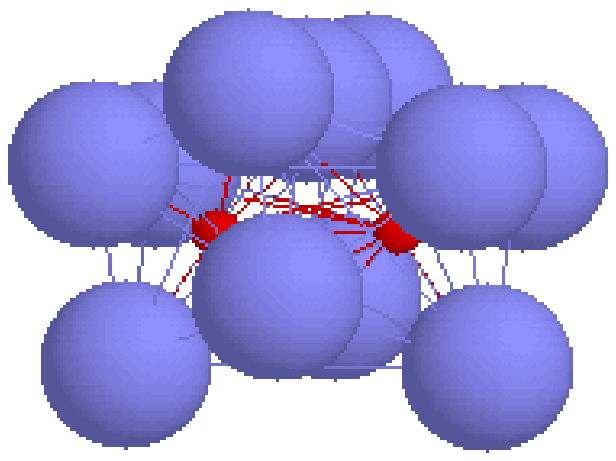} \\
    (8.1) & (8.2) \\
    \includegraphics[width=3.6cm,clip=true]{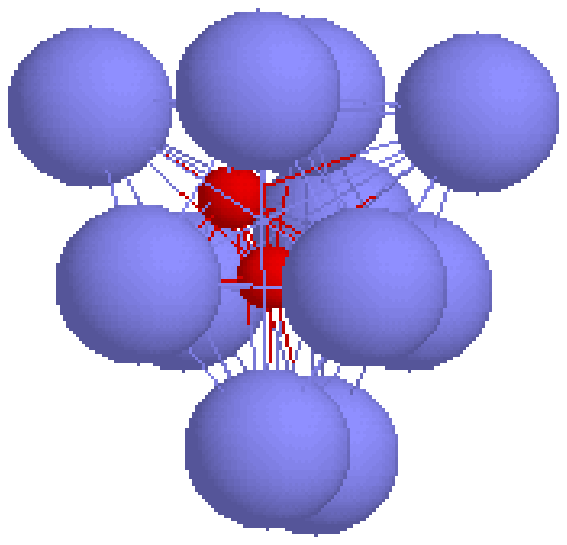} &
    \includegraphics[width=4.4cm,clip=true]{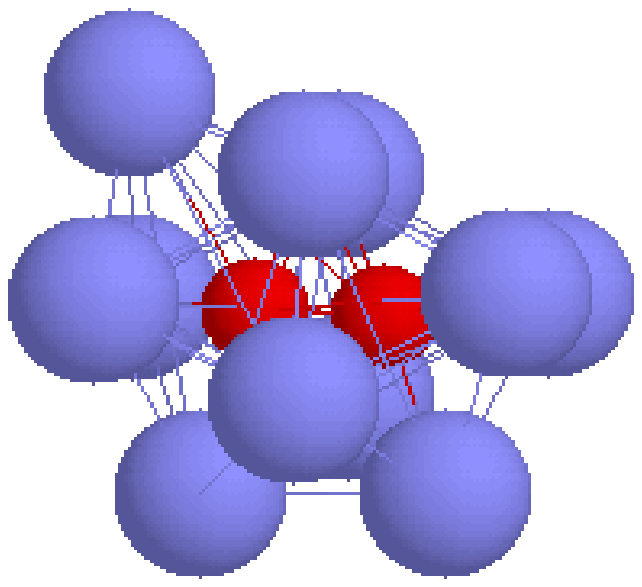} \\
    (8.4) & (8.6) \\
    \includegraphics[width=3.6cm,clip=true]{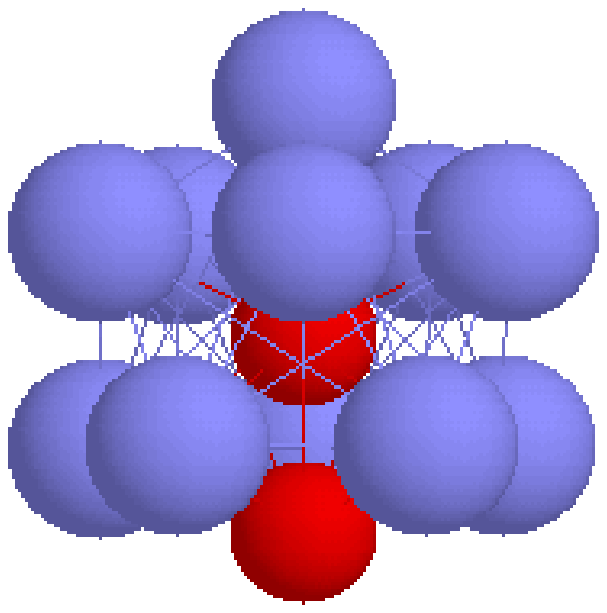} &
    \includegraphics[width=3.6cm,clip=true]{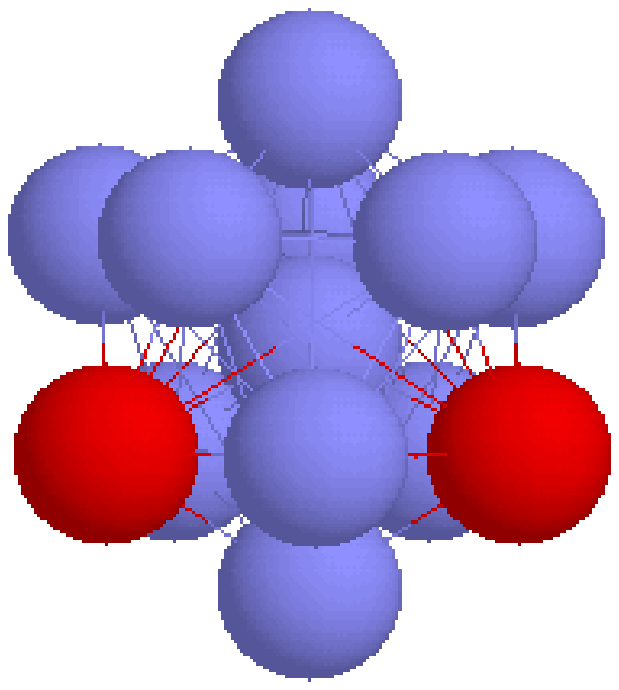} \\
    (8.8) & (8.9) \\
        \end{tabular}
\caption{\label{fig:coreEX11Y2}
Plots of selected $\mathrm{X_{11}Y_2}$ structures for various
$\mathrm(\sigma,\epsilon)$ values identified in Fig.~\ref{fig:corePESX11Y2}.
The number of the structures correspond to the regions labeled in
Fig.~\ref{fig:corePESX11Y2}. The core structures for
the systems labeled by (8.6) and (8.8) are not stable energy 
structures of the bare $\mathrm X_{11}$ system.}
\end{figure}

\begin{figure*}
  \begin{tabular}{@{}cc@{}}
    \includegraphics[width=8.5cm,clip=true]{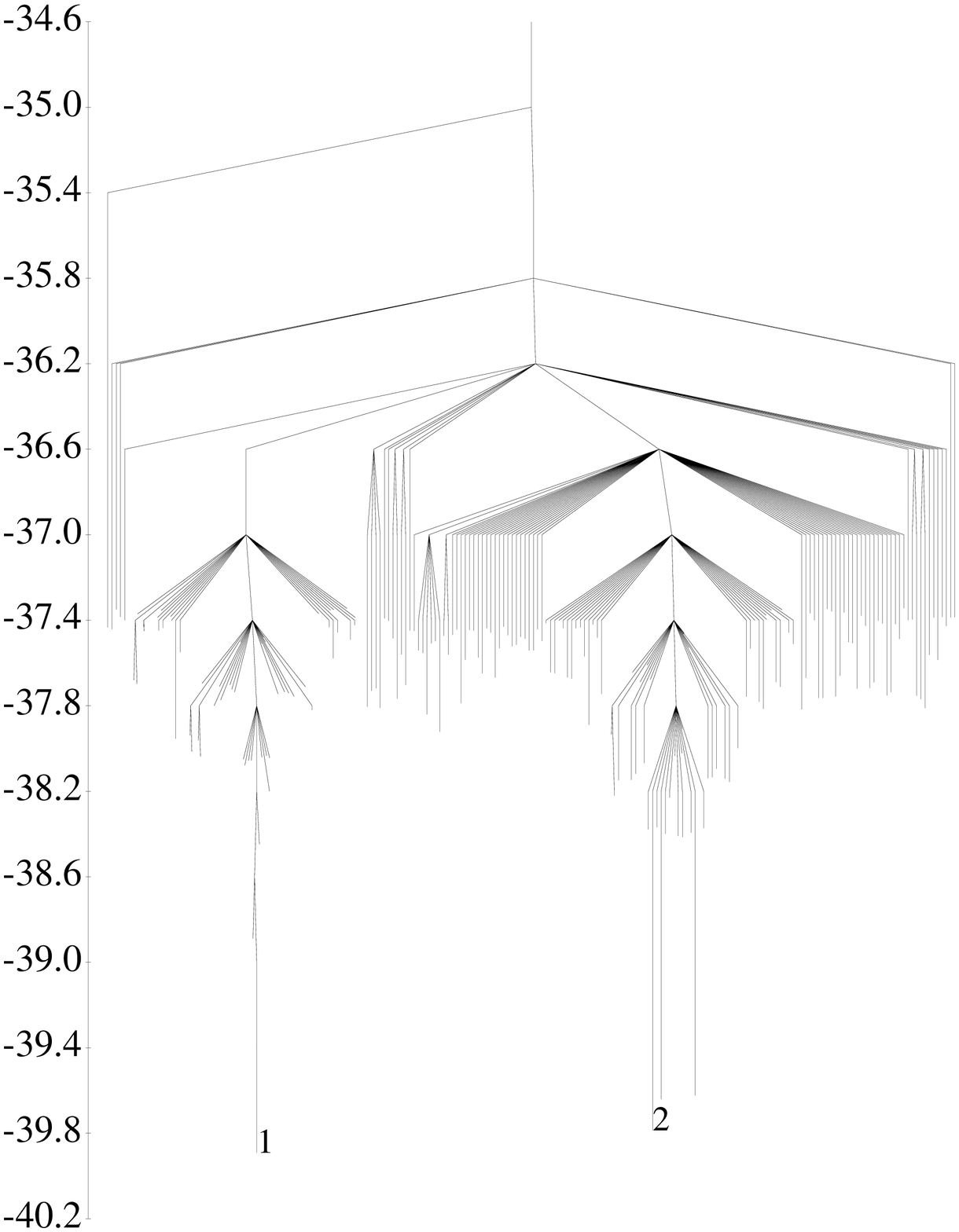} &
    \includegraphics[width=8.5cm,clip=true]{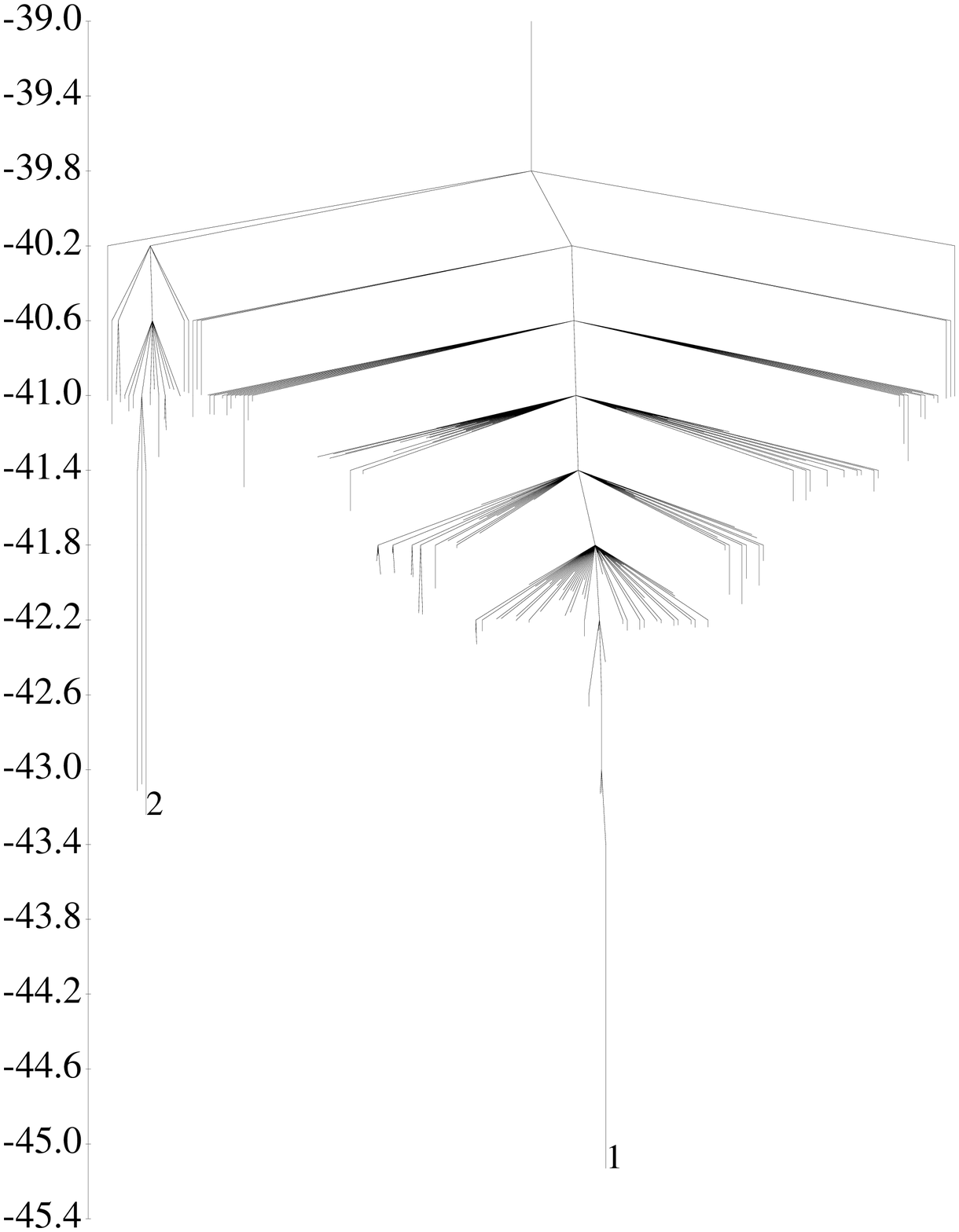} \\
    (a) & (b) \\
    \includegraphics[width=8.5cm,clip=true]{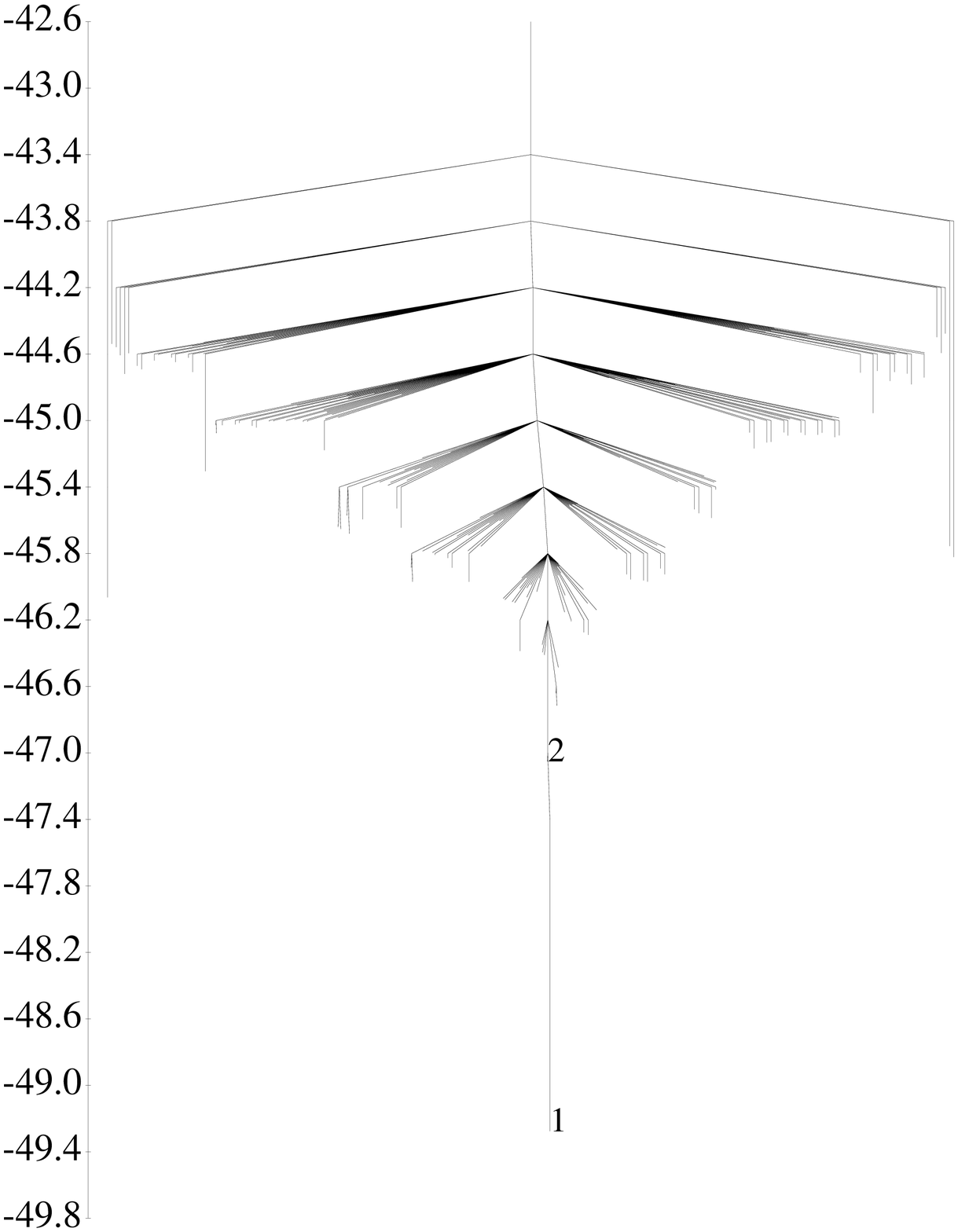} &
    \includegraphics[width=8.5cm,clip=true]{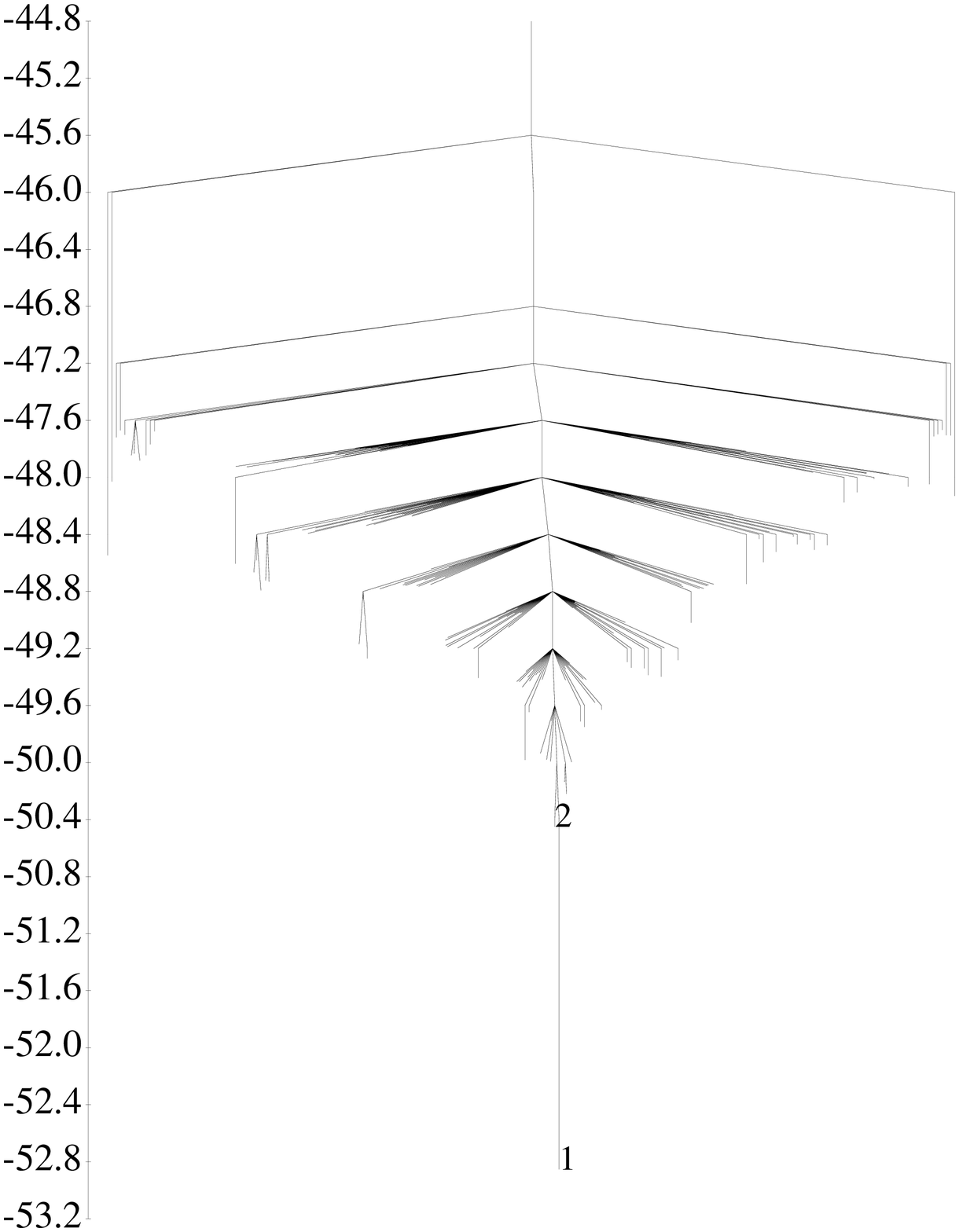} \\
    (c) & (d) \\
        \end{tabular}
\caption{\label{fig:DisConX11Y2}
Disconnectivity graph for $\mathrm{X_{11}Y_2(\sigma,\epsilon)}$
values demonstrating that we can control barriers for the selected
inherent structures. The energy scale is in units of $\epsilon_{XX}$.
The $\mathrm(\sigma,\epsilon)$ values for
panels (a--d) are (0.8,0.5), (0.8,1.0), (0.8,1.5) and (0.8,2.0),
respectively.
Only branches leading to the 200 lowest-energy minima are shown.}
\end{figure*}

As a second illustration, we consider mixed clusters of the type
$\mathrm{X_{11}Y_2}$. This system builds upon a parent, eleven-atom
system known to exhibit a set of 170,
energetically distinct inherent structures\cite{FAKEN01}.
The selected core inherent structures and associated energies
for the stable $\mathrm{X_{11}}$ inherent structures are presented in
Fig.~\ref{fig:core11}. In Fig.~\ref{fig:corePESX11Y2}, a 
$\mathrm(\sigma,\epsilon)$ contour plot of the core-atom potential energies
of the lowest total energy $\mathrm{X_{11}Y_2}$ clusters, again 
reveals the presence of definite ``core-phases''.
As illustrated in
Fig.~\ref{fig:coreEX11Y2}, some of these regions correspond to various
core structures present in the parent $\mathrm X_{11}$ system while others
correspond to new structures not seen in the original, single-component
cluster. We can see from Figs.~\ref{fig:corePESX11Y2} to 
\ref{fig:DisConX11Y2} that the impurity Y atoms provide us with significant
control over the relative ordering of the core energies of the parent
$\mathrm X_{11}$ system. Moreover, since we can manipulate the isomerization
barriers in the $\mathrm{X_{11}Y_2}$ systems, we can at least partially
stabilize clusters that exhibit selected core structures with respect to
isomerization. This is illustrated in Fig.~\ref{fig:DisConX11Y2}.

Figures~\ref{fig:DisConX11Y2}.a~--~\ref{fig:DisConX11Y2}.d represent the
$\mathrm{X_{11}Y_2}$ cluster at four points in Fig.~\ref{fig:corePESX11Y2}
with $\mathrm{X_{11}Y_2(\sigma,\epsilon)}$ coordinates (0.8,0.5), (0.8,1.0),
(0.8,1.5) and (0.8,2.0), respectively. The number of inherent structures
available to the $\mathrm{X_{11}Y_2}$ cluster in all four cases is more 
than 6000. We show only the lowest 200 inherent structures. The global minimum
of each system is labeled by the number 1 and contains as a recognizable
component the core structure shown in Fig.~\ref{fig:coreEX11Y2}.8.
In Fig.~\ref{fig:DisConX11Y2}.a the global minimum, the core structure
labeled by (8.8) in Fig.~\ref{fig:coreEX11Y2}, is linked to inherent
structure 5. Its core structure is different from the one associated
with the global minimum. The isomerization barrier between them is
$\Delta$E$_{1,5}$=1.015$\epsilon_{XX}$. In Fig.~\ref{fig:DisConX11Y2}.b
inherent structure 1 is connected to inherent structures 3 and 4
whose energies are almost degenerate. Both inherent structures 3 and 4
contain core structures that are different from each other and
from the one associated with inherent structure 1. The isomerization
barriers between inherent structures 1 and 3 and 1 and 4 are
$\Delta$E$_{1,3}$=2.025$\epsilon_{XX}$ and 
$\Delta$E$_{1,4}$=2.011$\epsilon_{XX}$, respectively.
Figures~\ref{fig:DisConX11Y2}.c and Fig.~\ref{fig:DisConX11Y2}.d show that
further increasing of the value of $\epsilon$ increases isomerization
barriers that link inherent structure 1 with inherent structure 2.
Numerically, these barriers are 
$\Delta$E$_{1,2}$=2.265$\epsilon_{XX}$ and
$\Delta$E$_{1,2}$=2.463$\epsilon_{XX}$, respectively.
Estimated lifetimes are shown in Fig.~\ref{fig:lifetimes}.b as a
function of $\epsilon$. It can be seen that the lifetimes increase
by fourteen and seven orders of magnitude at 5 and 10K, respectively.

\subsection{$\mathrm{X_{10}Y_3}$}

\begin{figure}[!htbp] \centering
  \begin{tabular}{@{}cc@{}}
    \includegraphics[width=3.6cm,clip=true]{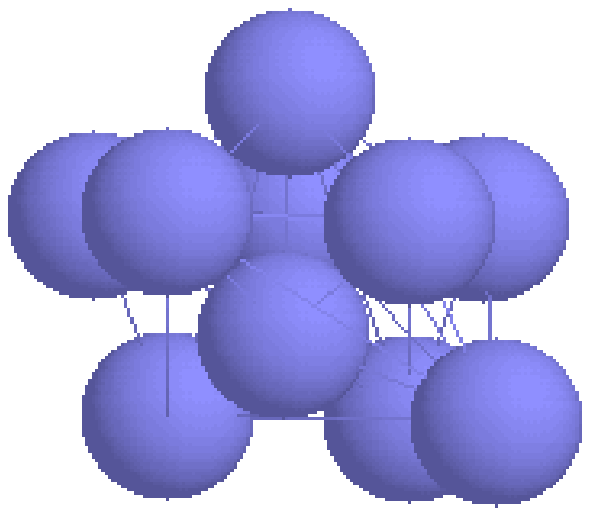} &
    \includegraphics[width=3.6cm,clip=true]{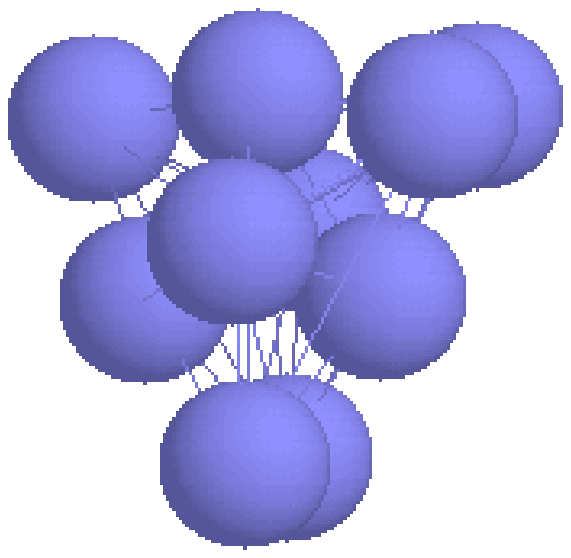} \\
    (a) & (b) \\
    \includegraphics[width=3.8cm,clip=true]{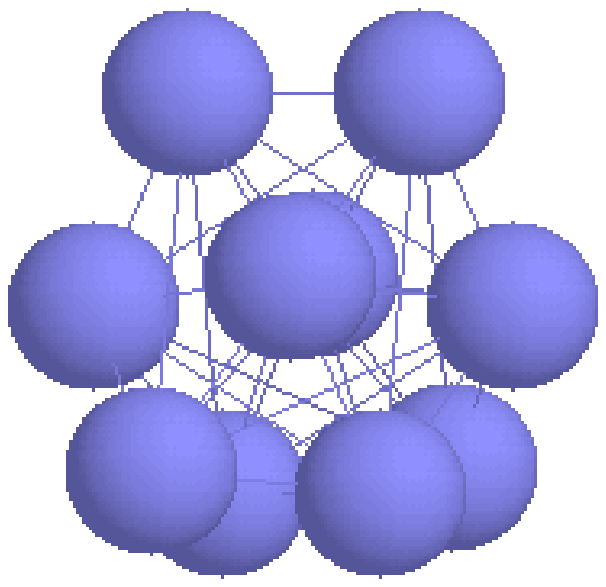} &
    \includegraphics[width=3.2cm,clip=true]{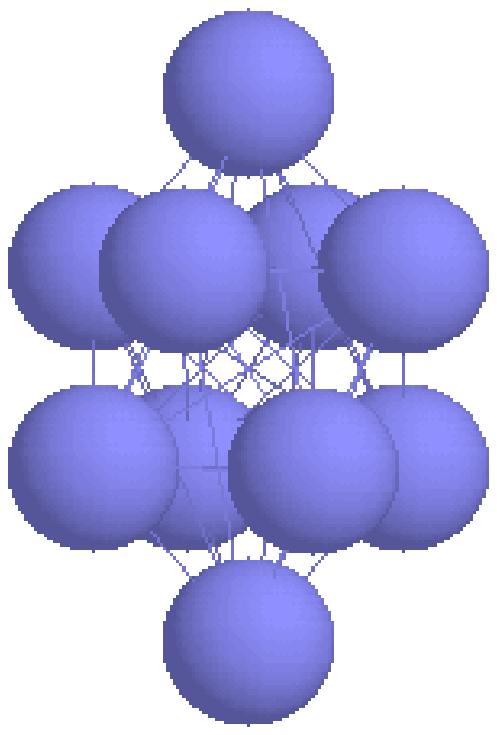} \\
    (c) & (d) \\
    \includegraphics[width=3.8cm,clip=true]{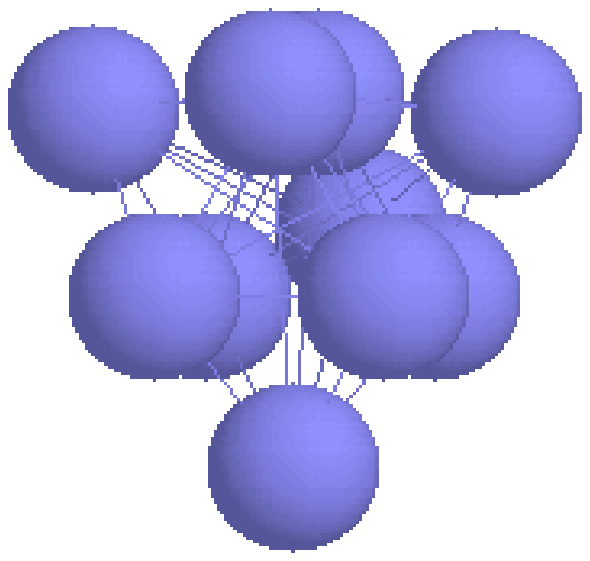} &
    \includegraphics[width=3.6cm,clip=true]{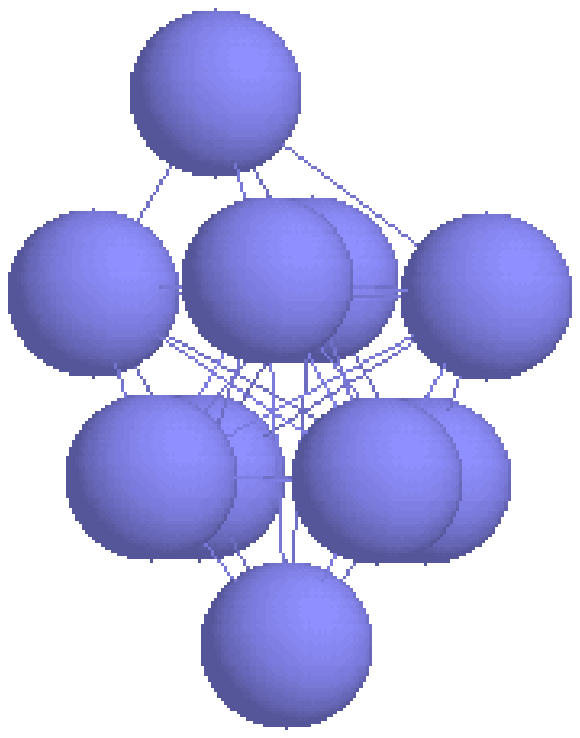} \\
    (e) & (f) \\
        \end{tabular}
\caption{\label{fig:core10} The selected stable inherent structure 
for X$_{10}$LJ cluster. 
Their energies (in units of the LJ well depth) are: (a) -28.422,
(b) -27.556, (c) -27.214, (d) -26.772, (e) -26.698, (f) -26.695.}
\end{figure}

\begin{figure}
\includegraphics[clip=true,width=8.5cm]{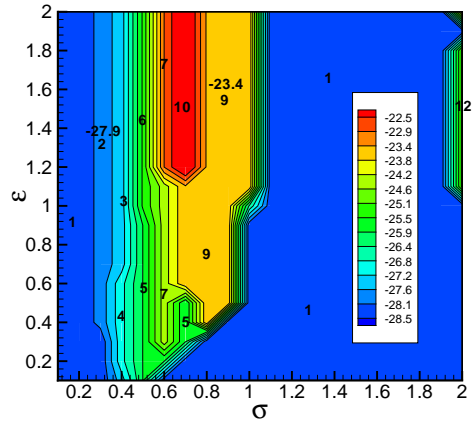}
\caption{\label{fig:corePESX10Y3}
$\mathrm E_{core}(\sigma,\epsilon)$ for $\mathrm{X_{10}Y_3}$.
Format for the plot is
the same as in Figs.~\ref{fig:corePESX12Y1} and ~\ref{fig:corePESX11Y2}.}
\end{figure}

\begin{figure}[!htbp] \centering
  \begin{tabular}{@{}cc@{}}
    \includegraphics[width=3.8cm,clip=true]{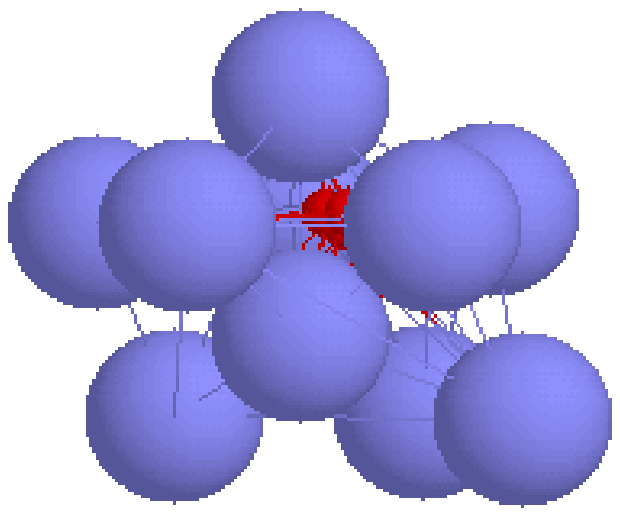} &
    \includegraphics[width=3.8cm,clip=true]{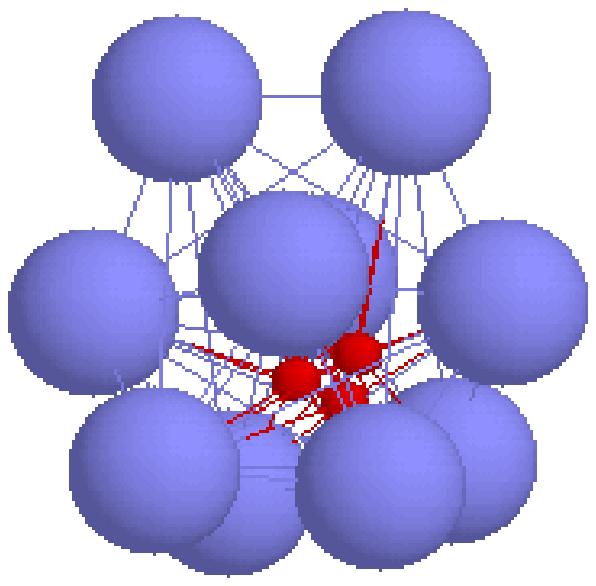} \\
    (12.1) & (12.2) \\
    \includegraphics[width=3.8cm,clip=true]{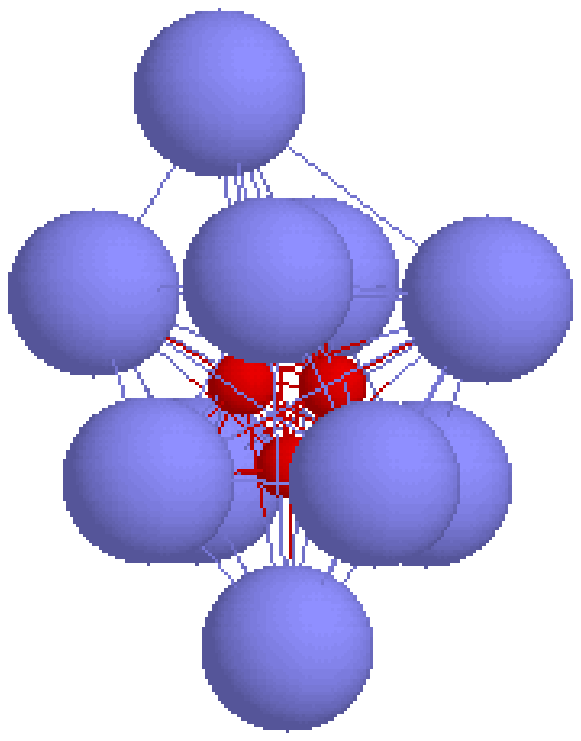} &
    \includegraphics[width=3.8cm,clip=true]{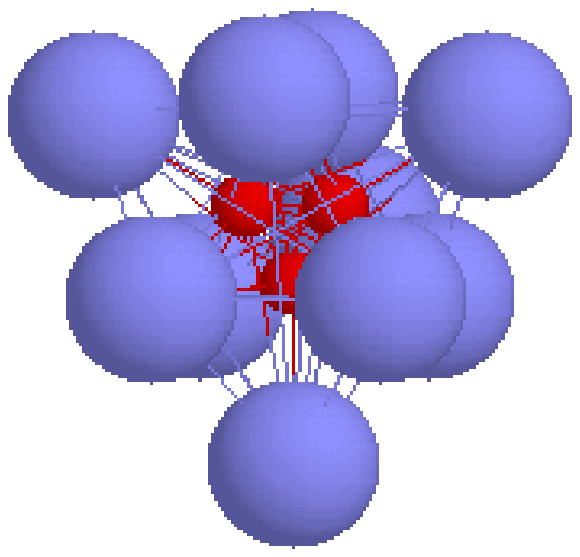} \\
    (12.3) & (12.4) \\
    \includegraphics[width=3.2cm,clip=true]{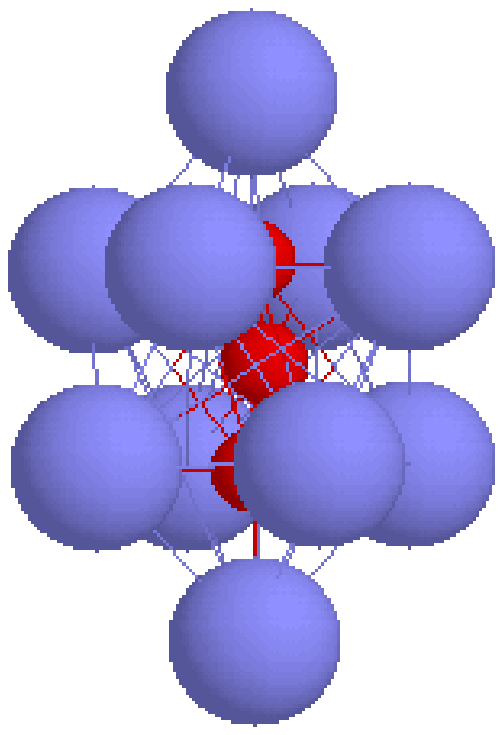} &
    \includegraphics[width=3.8cm,clip=true]{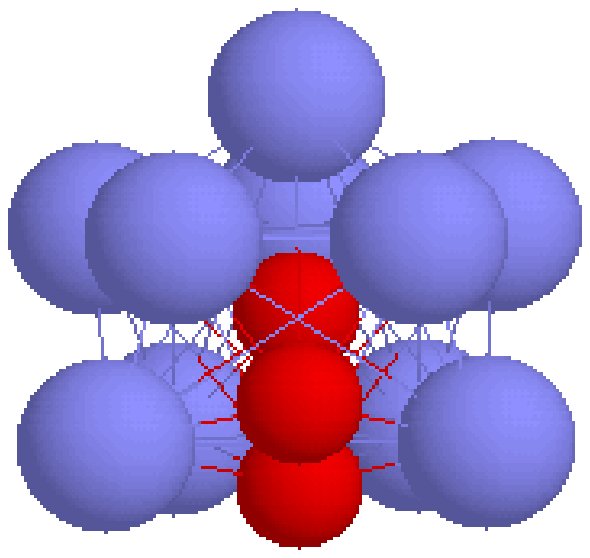} \\
    (12.5) & (12.9) \\
        \end{tabular}
\caption{\label{fig:coreEX10Y3}
Plots of selected $\mathrm{X_{10}Y_3}$ structures for various
$\mathrm(\sigma,\epsilon)$ values identified in Fig.~\ref{fig:corePESX10Y3}.
The number of the structures correspond to the regions labeled in
Fig.~\ref{fig:corePESX10Y3}. The core structure for
the system labeled by (12.9) is not a stable energy
structure of the bare $\mathrm X_{10}$ system.}
\end{figure}

\begin{figure*}
  \begin{tabular}{@{}cc@{}}
    \includegraphics[width=8.5cm,clip=true]{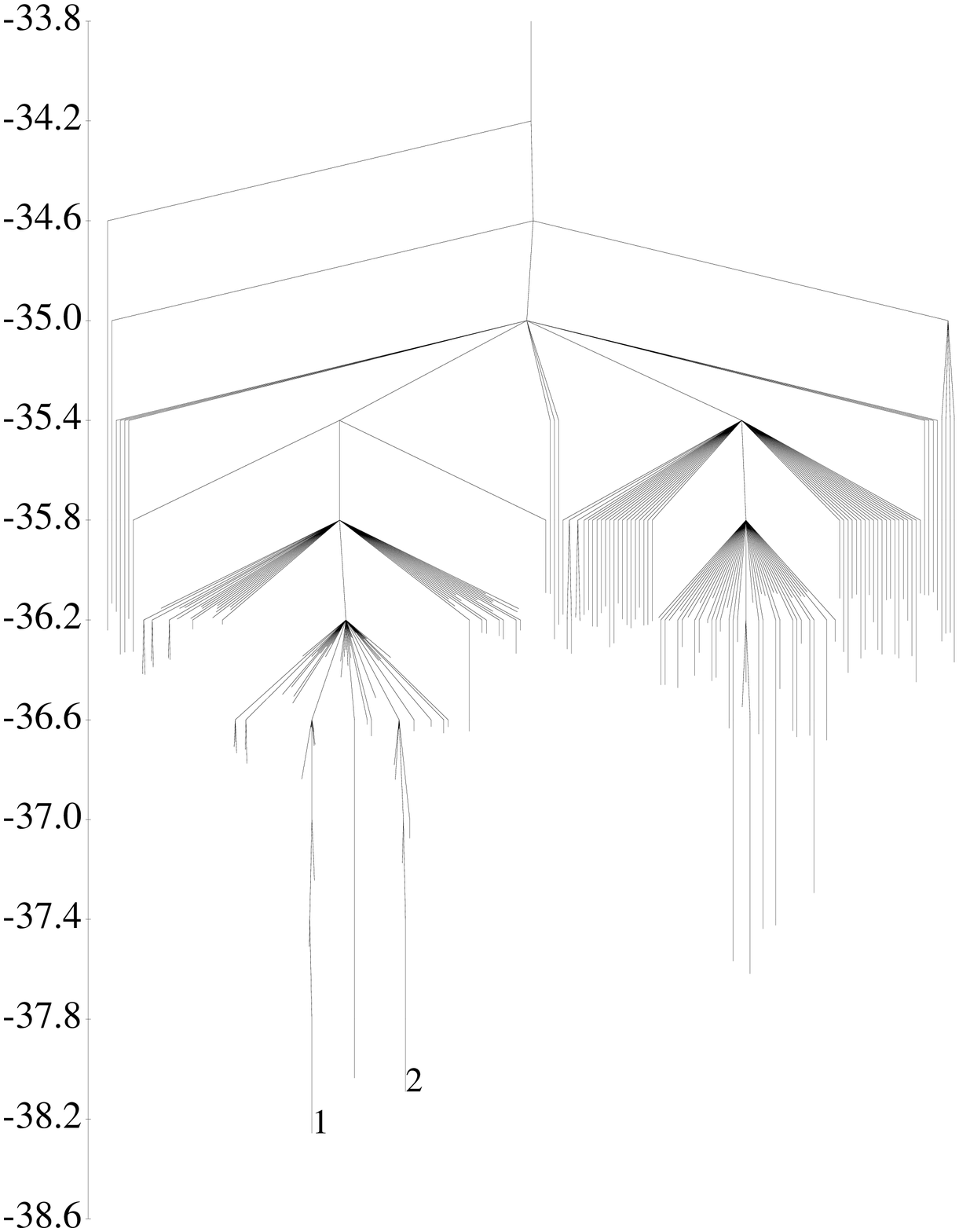} &
    \includegraphics[width=8.5cm,clip=true]{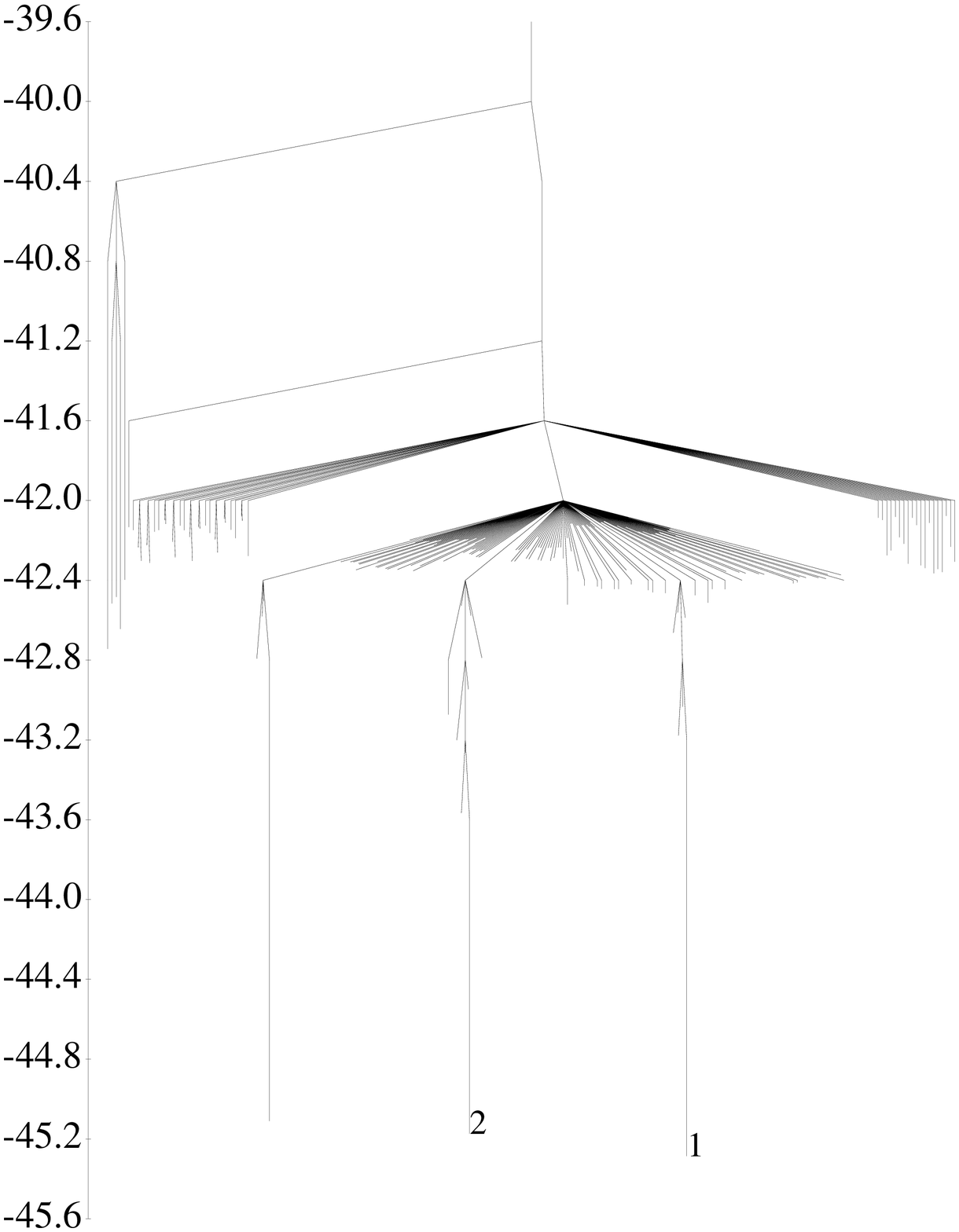} \\
    (a) & (b) \\
    \includegraphics[width=8.5cm,clip=true]{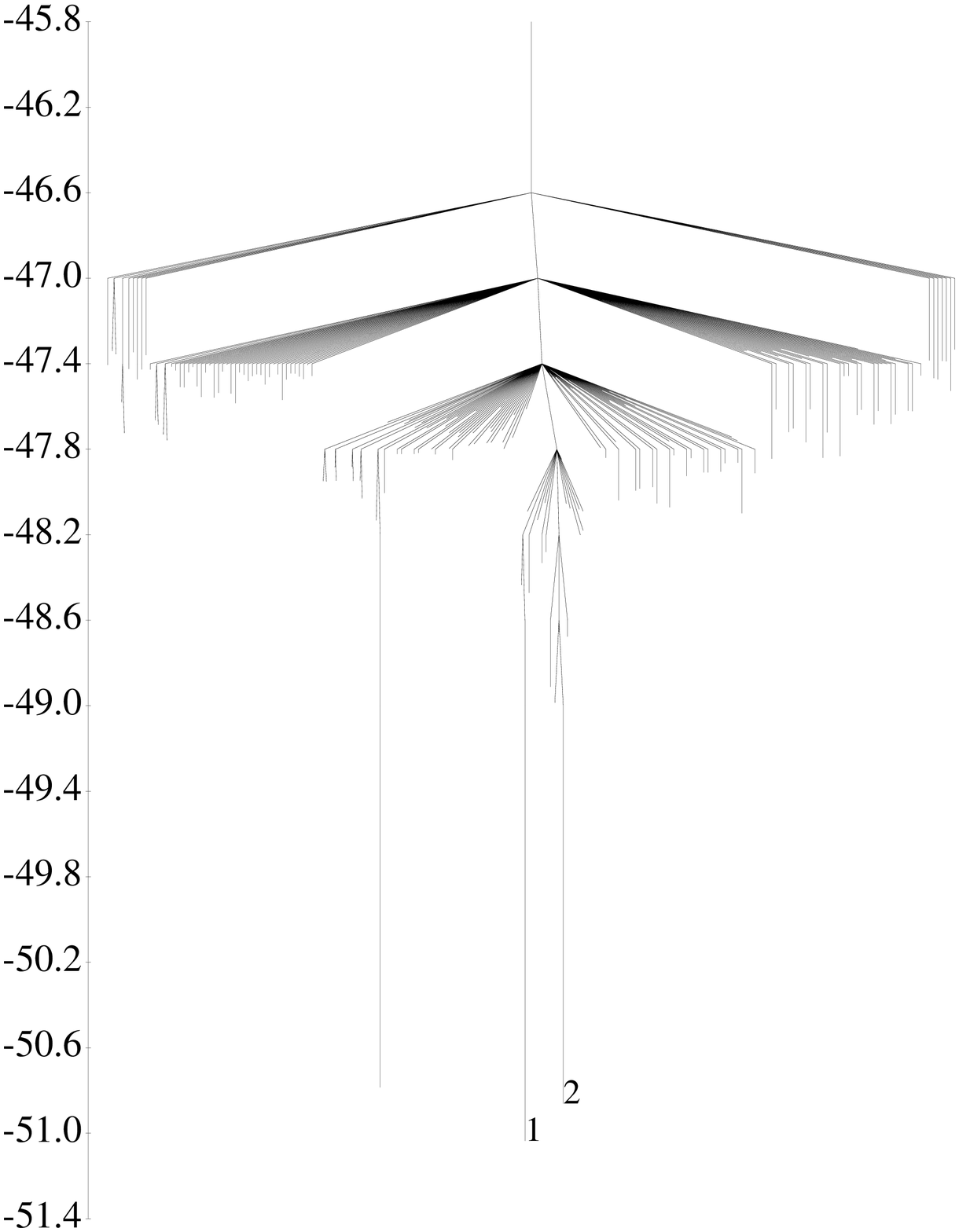} &
    \includegraphics[width=8.5cm,clip=true]{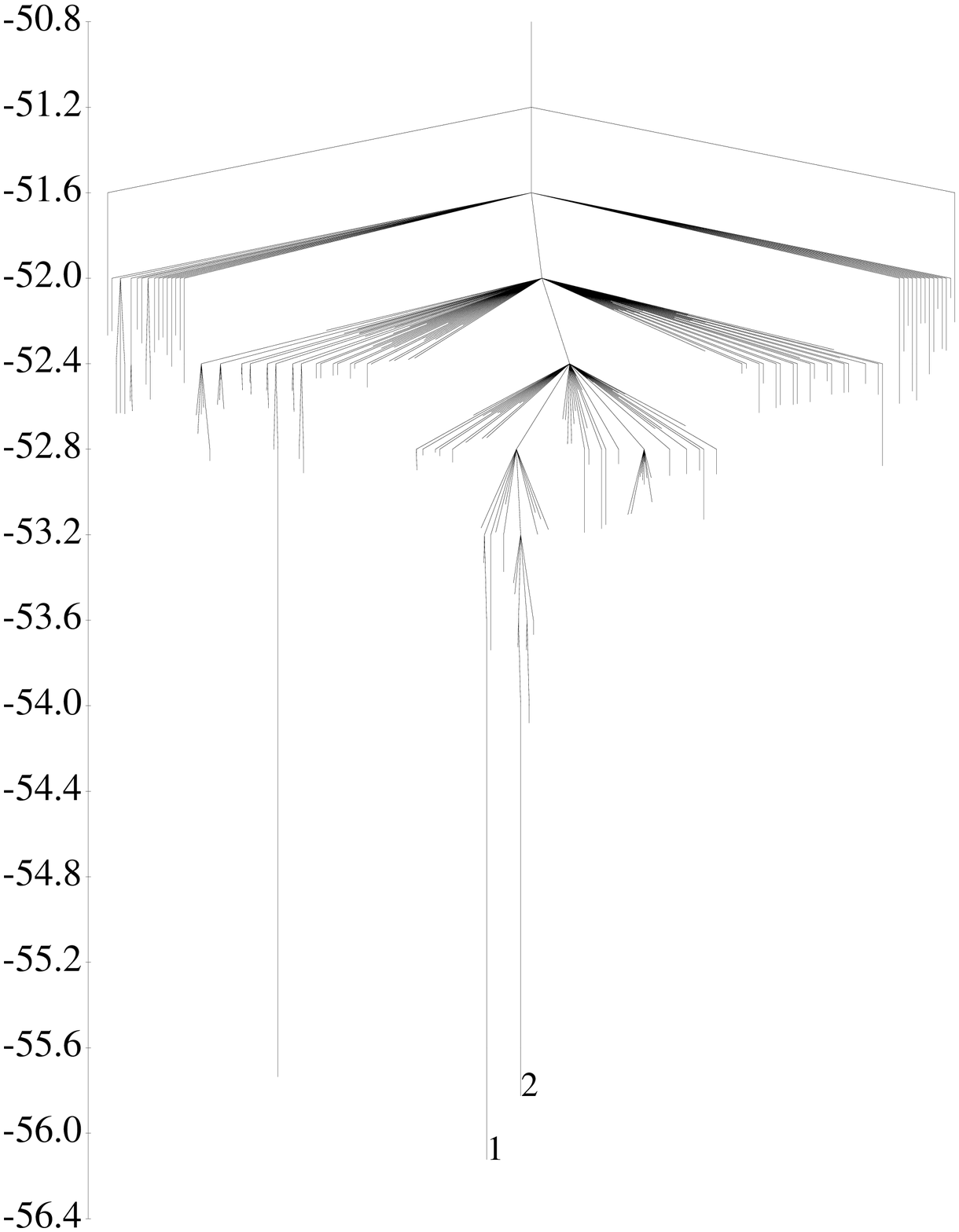} \\
    (c) & (d) \\
        \end{tabular}
\caption{\label{fig:DisConX10Y3}
Disconnectivity graph for $\mathrm{X_{10}Y_3(\sigma,\epsilon)}$
values demonstrating that we can control barriers for the selected
inherent structures. The energy scale is in units of $\epsilon_{XX}$.
The $\mathrm(\sigma,\epsilon)$ values for
panels (a--d) are (0.8,0.5), (0.8,1.0), (0.8,1.5) and (0.8,2.0),
respectively.
Only branches leading to the 200 lowest-energy minima are shown.}
\end{figure*}

As a third illustration, we consider mixed clusters of the type
$\mathrm{X_{10}Y_3}$. This system builds upon a parent, ten-atom, system
known to exhibit a set of 64, energetically distinct inherent
structures\cite{FAKEN01}. The selected core inherent
structures and associated energies for the stable $\mathrm{X_{10}}$
inherent structures are presented in Fig.~\ref{fig:core10}. A
$\mathrm(\sigma,\epsilon)$ contour plot of the core-atom potential
energies of the lowest total energy $\mathrm{X_{10}Y_3}$ clusters is
shown in Fig.~\ref{fig:corePESX10Y3}.
As illustrated in Fig.~\ref{fig:coreEX10Y3}
some of the domains in Fig.~\ref{fig:corePESX10Y3} correspond
to the core structures present in the parent system $\mathrm{X_{10}}$
while others correspond to new structures not seen in original, 
single-component cluster. Specifically, the domains 1, 2, 3, 4, and 5
correspond to the core structures labeled by (a), (c), (f), (e), and
(d) in Fig.~\ref{fig:core10}, respectively. Therefore,
the impurity atoms Y provide significant
control over relative ordering of the core energies
of the parent $\mathrm{X_{10}}$ system. Fig.\ref{fig:DisConX10Y3}
illustrates that, by choosing the appropriate set of $\mathrm(\sigma,\epsilon)$
values we can manipulate the isomerization barriers in the selected
systems.
\begin{figure}
\includegraphics[clip=true,width=8.5cm]{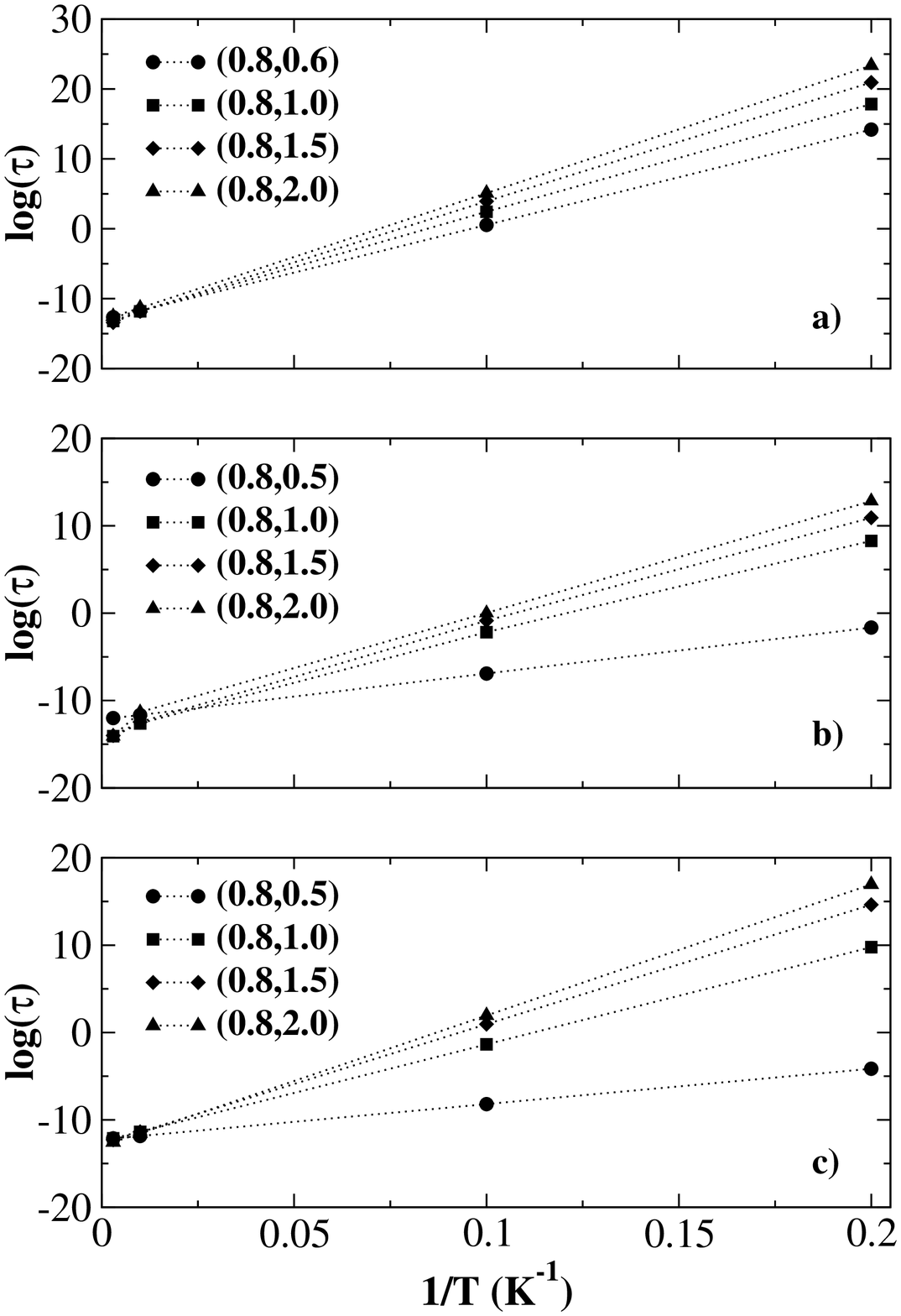}
\caption{\label{fig:lifetimes}
Temperature dependence of the logarithm (base 10) of lifetime $\tau$ 
(units seconds), for the selected inherent structures, as a function of 
different $\epsilon$ values. a) $\mathrm{X_{12}Y_1}(\sigma,\epsilon)$.
b) $\mathrm{X_{11}Y_2}(\sigma,\epsilon)$. 
c) $\mathrm{X_{10}Y_3}(\sigma,\epsilon)$.}
\end{figure}

Figures~\ref{fig:DisConX10Y3}.a~--~\ref{fig:DisConX10Y3}.d represent the
$\mathrm{X_{10}Y_3}$ cluster at four points in Fig.~\ref{fig:corePESX10Y3}
with $\mathrm{X_{10}Y_3(\sigma,\epsilon)}$ coordinates (0.8,0.5), (0.8,1.0),
(0.8,1.5) and (0.8,2.0), respectively. The number of inherent structures
available to the $\mathrm{X_{10}Y_3}$ cluster in all four cases is more
than 6000. We show only the lowest 200 inherent structures. The global minimum
of each system is labeled by the number 1 and contains as a recognizable
component the core structure shown in Fig.~\ref{fig:coreEX10Y3}.9.
In Fig.~\ref{fig:DisConX11Y2}.a inherent structure 1 is linked to 
inherent structure 6 with the isomerization barrier which value
is $\Delta$E$_{1,6}$=0.779$\epsilon_{XX}$. The core structure
corresponding to inherent structure 6 is different from the core
structure corresponding to the global minimum. In Fig.~\ref{fig:DisConX10Y3}.b
inherent structure 1 is connected to inherent structures 7 and 9 whose
energy values are -43.177$\epsilon_{XX}$ and 
-43.034$\epsilon_{XX}$, respectively. Both inherent structures 7 and 9
contain core structures that are different from each other and
from the one associated with inherent structure 1. The isomerization
barriers between inherent structures 1 and 7 and 1 and 9 are
$\Delta$E$_{1,7}$=2.138$\epsilon_{XX}$ and
$\Delta$E$_{1,9}$=2.252$\epsilon_{XX}$, respectively.
Figures~\ref{fig:DisConX11Y2}.c and Fig.~\ref{fig:DisConX11Y2}.d show that
increasing of the value of $\epsilon$ increases isomerization
barriers that link inherent structure 1 with inherent structures 9 and
12, respectively.
Numerically, these barriers are
$\Delta$E$_{1,9}$=2.630$\epsilon_{XX}$ and
$\Delta$E$_{1,12}$=2.883$\epsilon_{XX}$, respectively. 
It can be seen from Fig~\ref{fig:lifetimes}.c that by 
increasing the height of isomerization barriers the lifetime can
increase by twenty-one and ten orders of magnitude at 5 and 10 K,
respectively.
These results are specific demonstration of goal (3) stated earlier.

\section{Conclusions} \label{sec:conclude}

As stated at the outset, the general theme of the present work is to
explore the extent to which one can induce controllable structural
modifications in clusters. One reason for pursuing such a development
is the possibility that such modifications might be a general technique
for producing materials that have ``interesting'' properties (electronic,
magnetic, optical, thermal, etc.). Depending on the application, one
could envision such clusters being of use either directly, or, if they
could be made sufficiently robust, as precursors in subsequent assembly
of yet more complex materials.

Both current and previous work \cite{SABO03A} indicate that such
controllable modifications are possible in model Lennard-Jones systems.
In particular, we have demonstrated that by introducing impurity atoms
of varying size and interaction energies, we can produce core-atom
conformers that correspond to a variety of non-minimum energy homogeneous
isomers. We have also shown that it is also possible to use such impurities
to generate core-atom structures that posses no (stable) homogeneous
analogs. We have now demonstrated such capabilities for both the simple
$\mathrm{X_5Y_2}$ and $\mathrm{X_7Y_3}$ systems \cite{SABO03A} and for the
larger, more complex $\mathrm{X_{12}Y_1}$, $\mathrm{X_{11}Y_2}$, and
$\mathrm{X_{10}Y_3}$ binary clusters.

Finally, in the present work we have investigated the issue of the
stability of our modified clusters. Based on simple transition-state
estimates of the rates of isomerization of the various clusters, we have
demonstrated that it is possible both to induce and to stabilize a variety
of structural modifications. In the companion\cite{SABO04B} paper we
investigate how the structures of the underlying potential energy
surfaces explored in this work are reflected in the thermodynamic
properties of the systems.

{\bf Acknowledgment}

The authors acknowledge support from the National Science Foundation
through awards No. CHE-0095053. and CHE-0131114. They would also like
to thank Dr. M. Miller for helpful discussions and for his gracious
assistance with respect to the preparation of the disconnectivity graphs
in the present paper. Finally, the authors would like to thank the 
Center for Advanced Scientific Computing and Visualization at Brown
University for their assistance.

\bibliography{/aux/sabo/revtex4/revtex4/papers/sabo}


\end{document}